\documentclass[10pt,journal]{IEEEtran}

\usepackage{enumitem}
\usepackage{subcaption}
\usepackage{color}

\usepackage{balance}
\usepackage{csquotes}
\usepackage[normalem]{ulem}

\usepackage{hyperref}
\hypersetup{
 unicode=true,
 pdfstartview={FitH},
 colorlinks=true,
 citecolor=blue,
 linkcolor=blue,
 urlcolor=blue
}
\usepackage{pgfplots}
\usepgfplotslibrary{external}
\pgfplotsset{compat=1.16}

\usepackage{fancyhdr}
\usepackage{makecell}
\usepackage{rotating}
\usepackage{multirow}
\usepackage{booktabs}
\usepackage{tabularx}
\newcolumntype{L}[1]{>{\raggedright\let\newline\\\arraybackslash\hspace{0pt}}m{#1}}
\newcolumntype{C}[1]{>{\centering\let\newline\\\arraybackslash\hspace{0pt}}m{#1}}
\newcolumntype{R}[1]{>{\raggedleft\let\newline\\\arraybackslash\hspace{0pt}}m{#1}}

\usepackage{etoolbox}
\patchcmd{\quote}{\rightmargin}{\leftmargin 2em \rightmargin}{}{}

\usepackage{listings}
\newcommand{\listingcaption}[1]%
{%
\refstepcounter{lstlisting}{\hspace{-15pt} Listing \thelstlisting: #1}
}%
\newlength{\MaxSizeOfLineNumbers}%
\usepackage[ruled]{algorithm2e}

\SetAlFnt{\small}
\SetAlCapFnt{\small}
\SetAlCapNameFnt{\small}
\SetAlCapHSkip{0pt}
\IncMargin{-\parindent}
\usepackage{algorithmicx}

\usepackage{tikz}

\usepackage{cleveref}

\begin{document}

\markboth{ACCEPTED IN IEEE TRANSACTIONS ON COMPUTER-AIDED DESIGN OF INTEGRATED CIRCUITS AND SYSTEMS}{C. Pilato \MakeLowercase{\textit{et
al.}}: Optimizing the Use of Behavioral Locking for High-Level Synthesis}

\title{Optimizing the Use of Behavioral Locking for High-Level Synthesis}

\author{
Christian Pilato,~\IEEEmembership{Senior~Member,~IEEE,}
Luca~Collini,
Luca~Cassano,~\IEEEmembership{Member,~IEEE,}\\
Donatella~Sciuto,~\IEEEmembership{Fellow,~IEEE,}
Siddharth~Garg,~\IEEEmembership{Member,~IEEE,}
and Ramesh~Karri,~\IEEEmembership{Fellow,~IEEE}%
\thanks{Manuscript received November 1, 2021; revised January 31, 2022 and March 27, 2021; accepted May 10, 2022. This paper was recommended by Associate Editor Z. Zhang.

C. Pilato, L. Cassano, and D. Sciuto are with the Dipartimento di Elettronica, Informazione e Bioingegneria, Politecnico di Milano, Milano, Italy (contact email: christian.pilato@polimi.it).

L. Collini was with the Dipartimento di Elettronica, Informazione e Bioingegneria, Politecnico di Milano and is now with New York University, New York, NY, USA.

R. Karri and S. Garg are with the NYU Center for Cybersecurity 
(\url{http://cyber.nyu.edu}), New York University, New York, NY, USA.
}}

\maketitle

\begin{abstract}
The globalization of the electronics supply chain requires effective methods to thwart reverse engineering and IP theft. Logic locking is a promising solution, but there are many open concerns. First, even when applied at a higher level of abstraction, locking may result in significant overhead without improving the security metric. Second, optimizing a security metric is application-dependent and designers must evaluate and compare alternative solutions. 
We propose a meta-framework to optimize the use of behavioral locking during the high-level synthesis (HLS) of IP cores. Our method operates on chip's specification (before HLS) and it is compatible with all HLS tools, complementing industrial EDA flows. Our meta-framework supports different strategies to explore the design space and to select points to be locked automatically. We evaluated our method on the optimization of differential entropy, achieving better results than random or topological locking: 1) we always identify a valid solution that optimizes the security metric, while topological and random locking can generate unfeasible solutions; 2) we minimize the number of bits used for locking up to more than 90\% (requiring smaller tamper-proof memories); 3) we make better use of hardware resources since we obtain similar overheads but with higher security metric.
\end{abstract}


\section{Introduction}\label{sec:intro}

\begin{figure}[t]
\begin{subfigure}{\columnwidth}
\centering
\includegraphics[width=\columnwidth]{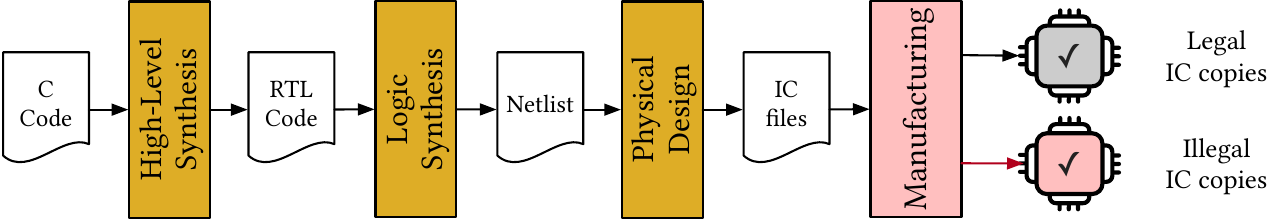}
\caption{Traditional HLS-based IC design flow.}
\label{fig:design_flow}	
\end{subfigure}
\begin{subfigure}{\columnwidth}
\centering
\includegraphics[width=\columnwidth]{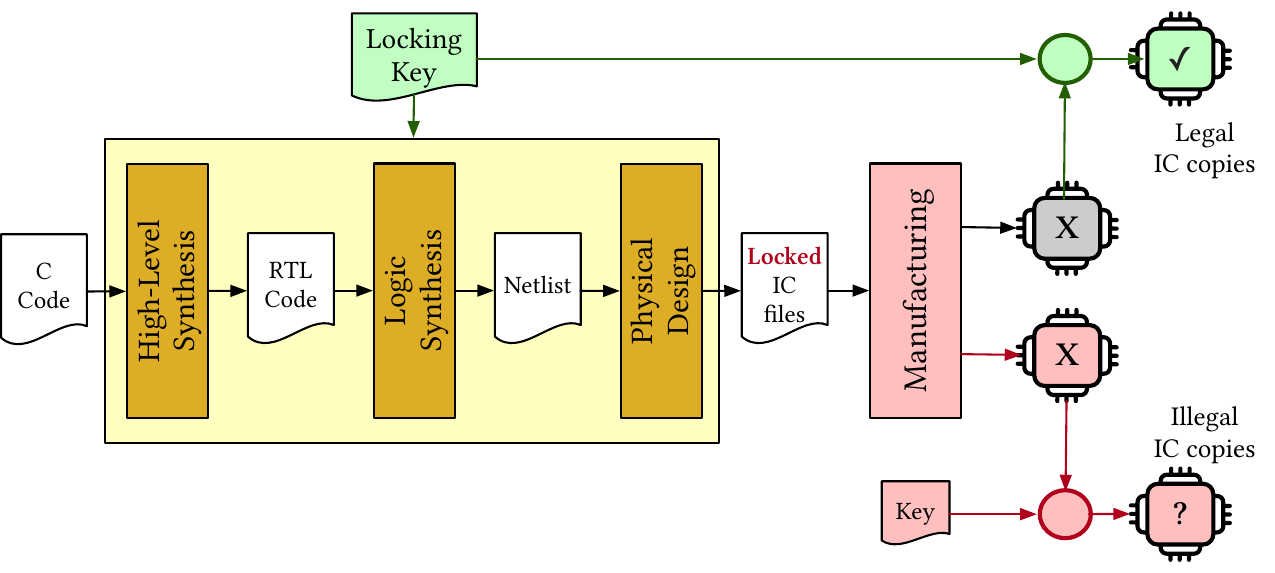}
\caption{IC design flow extended with locking.}
\label{fig:HLS}	
\end{subfigure}
\caption{HLS design flow. Red elements are untrusted entities.\vspace{-1.5em}}
\end{figure}

Due to the end of Dennard scaling, modern System-on-Chip (SoC) architectures are increasingly complex and heterogeneous, integrating several processor cores, memories, and specialized hardware accelerators~\cite{horowitz_isscc_2014}. Such complexity is pushing the design of integrated circuits (ICs) towards system-level methods based on \textit{high-level synthesis} (HLS)~\cite{nane2015survey}. \Cref{fig:design_flow} shows an example of HLS-based IC design flow, where the designers use HLS tools to automatically translate high-level, C-based specifications into register-transfer level (RTL) descriptions. Logic and physical synthesis generate the layout files ready for fabrication. HLS allows designers to raise the abstraction level, focusing on the behavior rather than hardware details and significantly improving design productivity. 

At the same time, IC's manufacturing costs are growing. For example, the equipment becomes 5$\times$ more expensive when scaling from 90nm to 7nm~\cite{iccost}. Many semiconductor companies cannot afford these costs and are becoming {\em fab-less}, outsourcing the IC fabrication to third-party foundries. This process creates security concerns~\cite{fabless}. Since the foundry has access to the design files, a rogue employee can analyze them to steal the intellectual property (IP) and create illegal IC copies~\cite{pieee14}. Design companies are using several techniques to thwart reverse engineering and IP counterfeiting~\cite{10.5555/2756806}.

Logic locking is a well-known technique for IP protection~\cite{6241494}. A high-level view of locking-aware design flow is depicted in \Cref{fig:HLS}. At design time,  gates are added to hide the correct function. These gates are controlled through an additional input signal ({\em locking key}) that is known to the design house but not to the foundry. After fabrication, the design house can \textit{activate} the correct IC function by placing the locking key in a tamper-proof key-storage element~\cite{RAHMAN202039}. 
This process can apply at different abstractions and assumes the attacker does not have and cannot guess the locking key. An IC with an incorrect key produces wrong results. On the other hand, the attackers should not be able to determine which results are clearly wrong to rule out incorrect keys and reduce the search space.
While locking has been widely studied,  many open issues remain~\cite{10.1145/3342099}. First, it must provide sufficient security protection from structural and functional viewpoints without suggesting to the attacker which keys are clearly wrong. Second, the cost should be minimized~\cite{8465830}. 
Third, the technology of the key-storage can limit the number of key bits that can be used.
However, the effects of locking depend on the chip function and are difficult to be predicted. 

{\em Behavioral locking} addresses these concerns by locking a design at a higher level of abstraction~\cite{8465830}. Behavior locking methods operate  at or above RTL and allow designers to protect semantic information before it is optimized and embedded into the netlist by logic synthesis. These methods scale to larger designs by reasoning about the design behavior instead of netlist structure. Their industrial adoption is limited since they require custom HLS tools. A valid alternative is to operate at the specification level (e.g., the input C code)~\cite{8715083},  assuming that HLS preserves the behavior, including the locking effects. However, in both cases designers miss a method to select which elements to lock, incurring overheads and producing weak or infeasible solutions~\cite{8465830}. 

This work follows the key idea that \textbf{locking all elements of a design does not necessarily provide maximum security}. The effects of some locking transformations may have limited visibility on the outputs or can be partially cancelled out by other transformations. The optimization process is application dependent and requires design space exploration. The selection should be guided by the analysis of the effects on the security metric. A designer must explore the application of locking transformations to identify the combination that maximizes the security metric while limiting the resource overhead. So, {\em \uline{optimizing a security metric requires a complex design space exploration that depends on the effects of the locking transformations on the design function}}. 

We propose a design framework to explore the functional effects of existing locking techniques at the C level and optimize their use. Our main contributions are: 
\begin{itemize}
\item a \textit{meta-framework} that integrates state-of-the-art C-level locking which allows use of HLS tools to generate locked RTL, enabling integration into IC design flows. 
\item A design-space exploration strategy (solution encoding and meta-heuristics) to select the best combination of locking points to optimize given security metrics.
\item A proof-of-concept implementation for the optimization of different entropy with a standard genetic algorithm.
\end{itemize}
Since our framework uses a standard integer-based encoding of the solutions, the designer has the possibility to integrate any state-of-the-art exploration algorithms.

The rest of the paper continues as follows. After introducing the threat model and motivating the work (Section~\ref{sec:background}), we present our {\bf design framework} to apply behavioral   locking with the support of commercial HLS (Section~\ref{sec:framework}). In this section, we also detail the different components: solution representation and analysis (\Cref{sec:analysis}), design space exploration (\Cref{sec:dse}), and solution evaluation (\Cref{sec:solution_eval}). Finally, we present a {\bf proof-of-concept implementation and evaluation} of our approach (Section~\ref{sec:results}).

\section{Problem Definition}\label{sec:background}

In the following, we show that identifying the points to be lock is a complex and application-dependent problem that requires to explore the design space. 

\vspace{4pt}
{\bf Problem Formulation:}
{\em Given a C specification and a locking key {\em K}, select the design points to be locked along with the corresponding parameters, such that the corresponding RTL solution has two properties: 1) it is one of the solutions with the best security metric; 2) it requires the minimal amount of hardware resources compared to other solutions. }

\vspace{4pt}
\noindent This problem formulation has the optimization of the security metric as the \textit{primary goal}, determining the design with minimal resources only afterwards. In the rest of this section, we define the threat model and the security metric (along with a motivating example) that we consider for the proof-of-concept implementation in this work.

\subsection{Threat Model}

We base our work on existing solutions for behavioral and RTL locking~\cite{8465830,pilato2021assure}. These methods assume {\em an untrusted foundry} that wants to identify the functionality of the given IC, i.e., the correct RTL implementation of the IP and its behavior over time, to make illegal IC copies. The untrusted foundry has access to the layout files of the locked chip. From these files, the foundry can reverse engineer the types of modules used in the design (i.e., registers, functional units, interconnection elements) and can identify the operations executed by each functional unit~\cite{7100906}. With this RTL description, the foundry can perform RTL simulations with different input and locking key values to extract information from the circuit that can help reconstruct the functionality. If successful, the foundry has the possibility of creating illegal copies of the IP. 

In this work, we assume the untrusted foundry has \emph{neither} access to the correct key \textit{nor} to a functioning unlocked IC (\textit{oracle}). This model is common for low-volume IC customers where the activated chips are used and available only in sensitive designs (e.g., US DoD). Even in consumer electronics, when the foundry is fabricating the chip for the first time, we can assume that an activated chip is not yet available~\cite{10.1145/3342099}. 

When no activated chip is available, SAT attacks are not possible and the attacker can only use random methods and the defender has to make all possible key-dependent variants equally plausible without leaking any additional information to the attacker~\cite{metrics2018}. Indeed, behavioral locking has been demonstrated to be able to thwart a wide range of attacks when the attacker has no access to an unlocked chip~\cite{pilato2021assure}. The resulting solutions can be then combined with scan-based methods to protect the key against oracle-based attacks~\cite{dac_assure_2021}.

\begin{figure}
\centering
\includegraphics[width=\columnwidth]{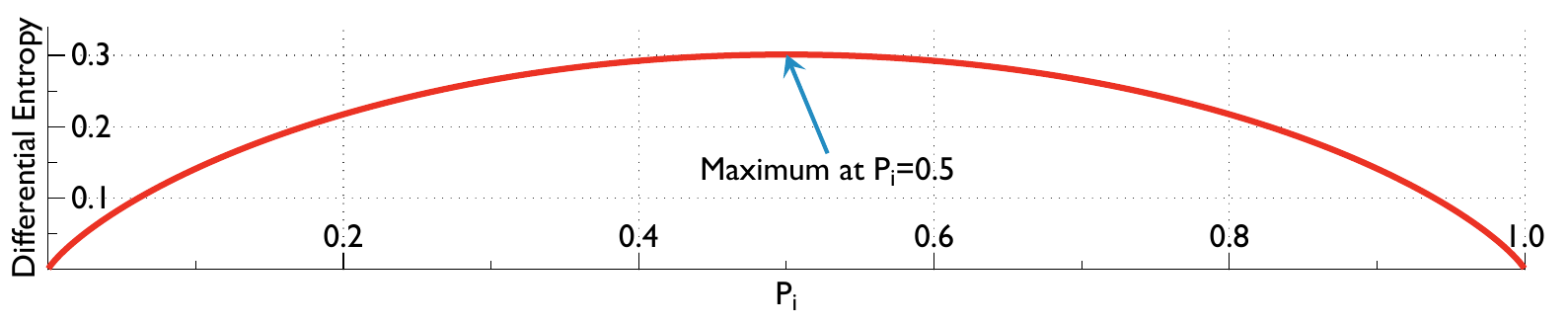}
\vspace{-16pt}\caption{Differential entropy for each value of $P_i$. The function has a maximum when $P_i=0.5$.}\label{fig:entropy}	
\end{figure}

\subsection{Locking Evaluation}\label{sec:metric}

Security evaluation is based on the assumption that only the chip activated with the correct key produces the expected results, while the other keys introduce errors making the corresponding chips unusable~\cite{pilato2021assure}. So, we evaluate the locking of a design $s$ based on the effects on the output results. Given  $N$ output bits, we compute the {\bf average differential entropy}~\cite{metrics2018} as follows:
\begin{equation}\label{eq:corruptibility}
H_s = \frac{sum_{i=1}^{N}\biggl(P^s_i \cdot log \frac{1}{P^s_i} + (1 - P^s_i) \cdot log\frac{1}{1-P^s_i} \biggr)}{N}
\end{equation}
where $P^s_i$ is the probability that the output bit $i$ of the locked design $s$ results different from its correct value. 
 The probability $P_i$ is estimated with random simulations, where different test cases (i.e., input sequences) and wrong key values are applied. Let $T$ and $W$ be the number of input sequences and wrong keys that have been provided for evaluation, respectively. The probability $P^s_i$ of each output bit $i$ is computed as:
\begin{equation}\label{eq:probability}
P^s_i = \frac{\sum_{w=1}^{W}\sum_{t=1}^{T} OUT[i]_t^g \oplus OUT[i]_{t,w}^s}{W\cdot T}
\end{equation}
where $OUT[i]_t^g$ represents the correct value of the output bit $i$ when the input sequence $t$ is tested, while $OUT[i]_{t,w}^s$ represents the actual value of the same output bit when the wrong key $w$ is provided to the given solution $s$ together with the same input sequence $t$.

The differential entropy metric is used to quantify {\em output corruptibility}, i.e. how much the locking techniques affect the outputs. This value should be maximized to avoid leaking any information on the correct output values to the attacker. Since $0 \leq P_i \leq 1$, Eq.~\ref{eq:corruptibility} has a maximum value $\hat{H_s}$ when $P_i = 0.5$ for each output bit~$i$ (see \Cref{fig:entropy}). This corresponds to the case where each output bit assumes value $0$ or $1$ with equal probability when wrong key values are applied. As a result, the attacker has no information on the correct output values and can only make random guesses. For this reason, our framework aims at maximizing $H_s$.
Although we used differential entropy, our methodology is general and requires a security metric to evaluate each candidate for the given threat model.

\subsection{Behavior Locking}\label{sec:mot}

\begin{figure}[t]
\centering
\includegraphics[width=0.90\columnwidth]{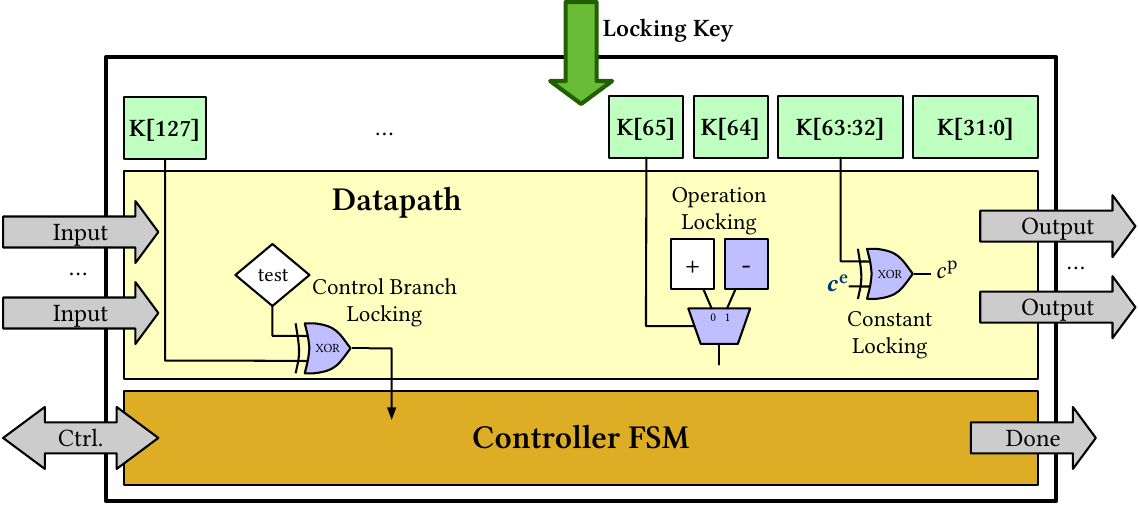}
\caption{HLS-generated IP with behavior locking.}\label{fig:lock_circuit}
\end{figure}

Behavioral locking hides parts of the function (e.g,. constants, control branches and arithmetic operations) based on the locking key $K$. It can be applied on C code~\cite{8715083}, during HLS~\cite{8465830}, or at RTL~\cite{pilato2021assure}. The key $K$ is provided by the designer through an input port and partitioned into sub-keys to lock each element, as shown in \Cref{fig:lock_circuit}. The circuit will work correctly only when the correct key is given. This approach is more scalable than gate-level locking, protecting the semantics of the design instead of its structural netlist. 
We consider the following behavior locking techniques~\cite{8465830,8715083,pilato2021assure}.

\vspace{4pt} 
{\bf Control Branch Locking.} Branches in the input behavior can be locked to hide the control flow. Each condition can be locked with one bit key. The condition $c_p == 1$ is modified as $c_p \oplus k_j == 1$, where $k_j$ is a one-bit key. This $k_j$ is part of the locking key 
$K$ and locks this condition checking. The required branch is taken only when the correct $k_j$ is provided.

\vspace{4pt} 
{\bf Operation Locking.} Fake operations are added to hide real RTL operations. Given an operation $l$ to be locked, the outputs of the two operations (correct and fake) are multiplexed by a key bit $k_l$. The correct output is connected to $0$ or $1$ input of this MUX based on the value of $k_l$. Only with the correct key, the correct operation results are produced.

\vspace{4pt} 
{\bf Constant Locking.} We assume a predefined number of bits $x$ to implement all constants, typically 32 bits (corresponding to an integer value in C), regardless the real bit-widths. Each constant $c_i^p$ of the behavior is locked as $c_i^e = c_i^p \oplus k_i$, where $c_i^e$ is the locked value stored in hardware and $k_i$ is a $x$-bit key. The correct constant can be obtained in hardware by reversing the operation, i.e., $c_i^p = c_i^e \oplus k_i$.  

\vspace{4pt}
We define a {\bf locking point} as any of the RTL elements (i.e., a control branch, an operation, or a constant) where it is possible to apply the given locking techniques. 

\subsection{Motivating Example}
While behavioral   locking is a powerful solution to hide the IC functionality, we argue that locking a large number of locking points may produce a large overhead~\cite{8465830} without necessarily improving the given security metric. 
Consider locking the  cyclic redundancy check (CRC) code IP. For simplicity, we use Bambu HLS  tool~\cite{bambu_fpl_2012} targeting a Xilinx Virtex-7 XC7VX690T FPGA at 100MHz. The algorithm has 5 operations and 7 constants that can be locked with 167 bits. When we constrain behavioral locking to use no more than 50\% of these bits and we use TAO approach~\cite{8465830} (this  is also known as \textit{topological locking}), the RTL  has an overhead of 1,430 look-up tables (LUT) and 815 flip-flops (FF) compared to the unlocked version. Differential entropy of the design is $\sim$50.53 (where maximum is 64) and the algorithm locks all operations and 2 constants. A high differential entropy (63.08) can be achieved by selecting and locking only 5 operations and 1 constant. This uses 730 LUTs and 385 FFs with a reduction of overhead by about 50\%. Thus optimizing behavioral  locking is important to improve security metric and reduce overhead. 

\section{Proposed Exploration Meta-Framework}\label{sec:framework}

\begin{figure}[t]
\centering
\includegraphics[width=\columnwidth]{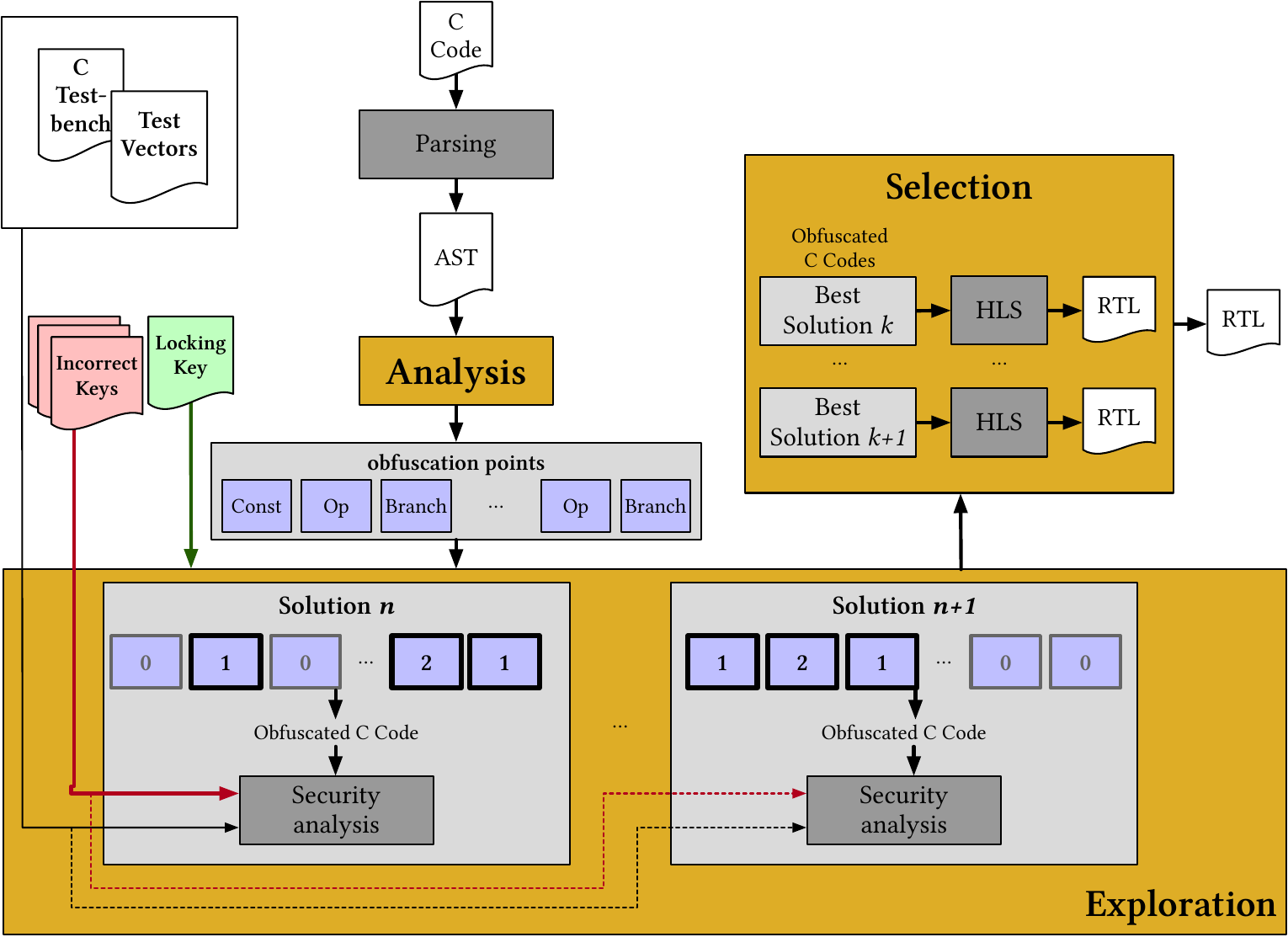}
\caption{Framework to optimize behavioral locking.}\label{fig:framework}	
\end{figure}

We propose a {\bf modular and integrated meta-framework} (see \Cref{fig:framework}) to optimize the use of behavioral locking during HLS. The input is a synthesizable C code of the accelerator. Behavioral locking is applied as a source-to-source transformation on such input C code. In this way, we can leverage existing HLS tools for generating the locked RTL description. We assume that the input C code is already synthesizable with the given HLS tool.
Since we consider a metric that analyzes only the IC behavior and we assume that HLS-generated designs have the same behavior of the corresponding input C codes, we can perform security assessment directly on the locked C code. This approach is much faster than performing RTL simulations, enabling its use in an exploration framework.
To perform locking, we provide the locking key $K$. The locking key must be independent of the design to avoid that the attacker can infer it. So, it is an input of our methodology. The size of the locking key determines the maximum number of bits that can be used for locking, limiting the locking techniques that can be applied. 
To compute the differential entropy of each candidate solution (see Section~\ref{sec:metric}), the framework requires the corresponding C-based test-bench and a set $T$ of representative inputs to evaluate the behavior of locked circuit versions with correct and incorrect keys. Such input vectors are the same that are used to evaluate the circuit functionality. An additional set $W$ of wrong keys is also provided. Each wrong key in $W$ has the same length as the locking key $K$ and is randomly generated by altering the correct key.
By leveraging the given HLS tool, the framework outputs an RTL description of the best locked solution that is ready for the front- and back-end synthesis steps. 

Our exploration framework operates as follows.
First, we execute the input C code on the set $T$ of representative inputs to compute the {\em golden outputs}. These values will be used to assess the effects of applying the candidate set of locking techniques to the input C code. We parse the input C code to build the corresponding {\em abstract syntax tree} (AST) of the functionality to be locked. Each AST node describes a construct occurring in the source code. This representation aids the next steps since it can be analyzed to identify locking points and edited to create alternative locked versions. The rest of the framework has three main steps.
\begin{enumerate}[nosep,leftmargin=15pt,labelwidth=*,align=left]
	\item {\bf analysis}: we analyze the input C description to identify the potential locking points. A locking point is an element of the algorithm (i.e., constant, operation, branch condition) that can be potentially locked based on the available techniques (see Section~\ref{sec:mot}).
	\item {\bf exploration}: we perform design space exploration to identify the sub-set of solutions that optimize the given security metric. Each solution represents a combination of decisions concerning how to apply the techniques to each locking point. The corresponding locked C codes are generated by applying the locking techniques specified in the solutions to evaluate the security metric.
	\item {\bf selection}: we apply HLS on the set of locked C codes produced in the previous step and we determine the cost of the resulting RTL designs to select the final solution, which is our {\em best design}.
\end{enumerate}
In the {\em exploration} phase, we can use several strategies, ranging from random changes (similar to approach used for logic locking in \cite{roy2008epic}) to complex meta-heuristics like genetic algorithms and simulated annealing. Meta-heuristics are based on the observation of natural behaviors. For example, {\em simulated annealing} (SA) is inspired by annealing in metallurgy to create perturbations and move around the design space and a {\em genetic algorithm} (GA) maintains a population of alternative solutions to be recombined. 
These algorithms perform well in the identification of sub-structures in the problem~\cite{5467335,294849}.

\subsection{Identification and Representation of Locking Points}\label{sec:analysis}

\begin{figure}
 \centering
\includegraphics[width=\columnwidth]{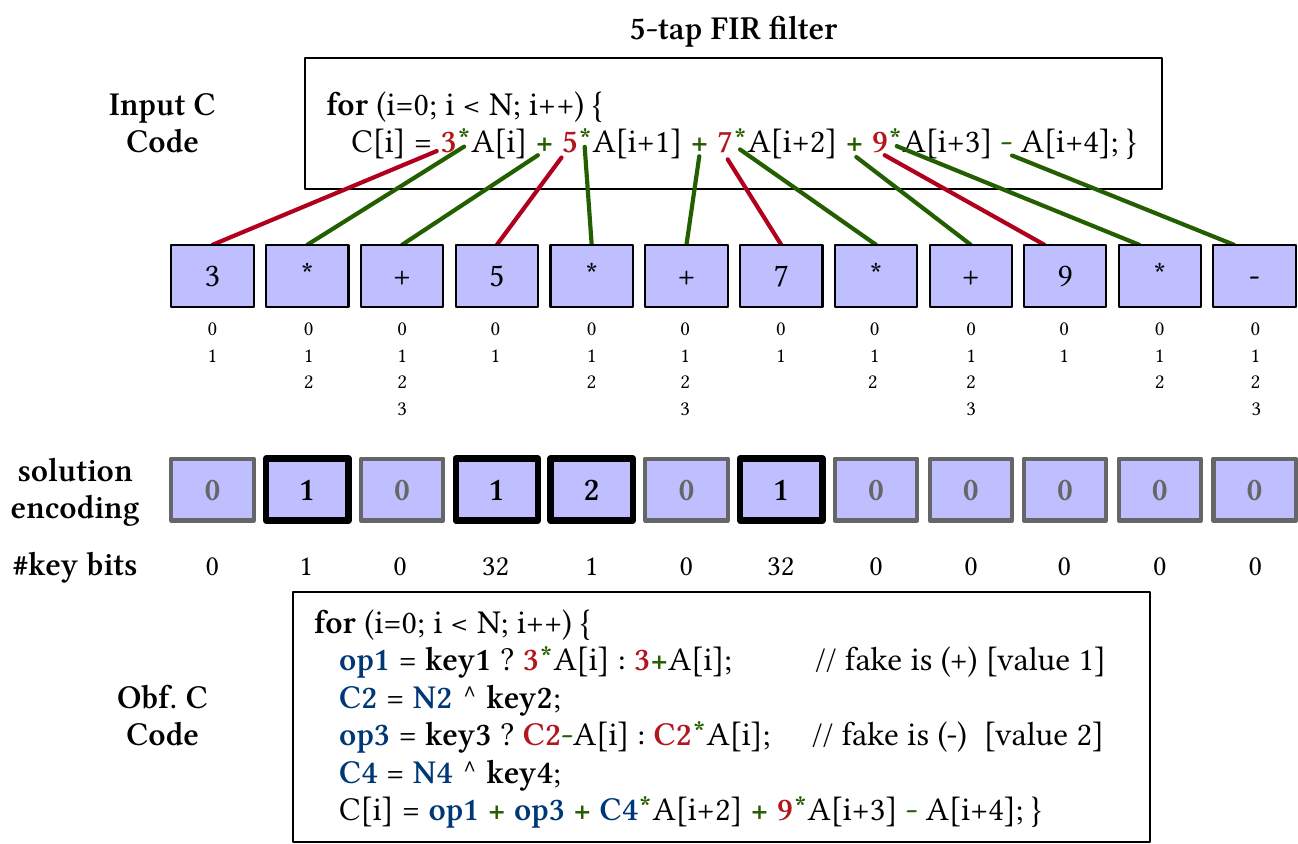}
\caption{Example of solution encoding. Note that N2 and N4 are the encrypted constants obtained by XOR-ing the original constant with the key (see $c_i^e$ in Section \ref{sec:mot})} \label{fig:encoding}
\end{figure}

During the {\em analysis step}, we perform a depth-first analysis of the AST of the input C code to identify the potential locking points in the design region to be protected. The type and number of locking points depend on the algorithm to be implemented and the locking techniques. We identify all potential locking points in the candidate region considering the techniques described in \Cref{sec:mot} as follows: 
\begin{itemize}[nosep,leftmargin=15pt,labelwidth=*,align=left]
\item {\bf constants}: we lock constants with a pre-defined number $B_c$ of key bits. The number of key bits is the same for all constants to prevent information leakage about the constant range. Each constant is represented  with {\tt 0}/{\tt 1} value to specify whether a constant value must be locked. Each constant to be locked (i.e., with the corresponding value set to {\tt 1}) requires $B_c$ key bits.  
\item {\bf operations}: we lock logic and arithmetic operations with extra fake operations. For each operation type, we pre-select a set of alternative types. Each operation $o$ is represented in the corresponding vector element with a value that ranges between {\tt 0} (no locking) and $N_o$, where $N_o$ is the number of alternative operation types pre-defined for the type of operation $o$. Each locked operation (i.e., with a value different from {\tt 0}) requires 1 key bit to multiplex the output of the correct operation with the fake one.
\item {\bf branches}: we lock the control flow (e.g., {\tt if}/{\tt else} statements or ternary operators) with key bits, reordering branches as needed. Each condition evaluation is represented in the solution with {\tt 0}/{\tt 1} value to specify whether the corresponding branch is locked or not. Each locked branch (i.e., value set to {\tt 1}) requires 1 key bit.
\end{itemize}
When a locking point $i$ has $O_i$ alternatives, the decision  can be represented with an integer value between $1$ and $O_i$ when it should be locked and $0$ otherwise (see the upper part of \Cref{fig:encoding}). For example, if an addition can be locked with two  types of ``fake'' operations: subtraction and multiplication, the corresponding element can take on: {\tt 0} (no locking), {\tt 1} (lock with subtraction), and {\tt 2} (lock with multiplication). On the contrary, a control branch can assume only two values: {\tt 0} (no locking), {\tt 1} (locking). So, the analysis creates a vector of integers that represents decisions for all locking points. 
\Cref{fig:encoding} shows a solution encoding for a simple algorithm. The integer vector represents a {\em locking solution} has as many elements as the number of locking points. It can be manipulated by meta-heuristics to generate alternatives and search the design space.

\vspace{4pt}
{\bf Key-bit Requirements.}
The  number of key bits $K^s$ required to lock a solution $s$ is:
\begin{equation}\label{eq:keys}
	K^s = \sum_1^{N^C} b_c * B_c + \sum_1^{N^O} b_o + \sum_1^{N^B} b_b
\end{equation}
where $N^C$, $N^O$, and $N^B$ is the total number of  locking points for constants, operations and branches, respectively. $b_c$, $b_o$, and $b_b$ have value {\tt 1} when the  value in the solution is different than {\tt 0} (i.e., the corresponding locking technique must be applied). $B_c$ is the  key bits pre-assigned to lock the constant $c$. This work considers $B_c = 32$ for all constants. The designer can  integrate additional constraints. Functions can be excluded from locking and from analysis. We can also force locking of specific parts, and explore how to spend the key bits. In this case, the {\em analysis} phase does not add value {\tt 0} to the list of values for the corresponding locking points, forcing them to have a value always different than zero. A solution $s$ is always valid since the locking techniques are orthogonal to each other, but it is {\em feasible} (i.e., it can be implemented in the target system) if and only if there are enough key bits in the key (i.e., $K^s \leq K$).

\vspace{4pt}
{\bf Size of the Design Space.}
The size of the design space corresponds to the combinations of locking techniques that the designer can apply. Given a  locking solution described by a vector of $N$ \textit{candidate locking points} ($N^C + N^O + N^B$), the number of different solutions is:
\begin{equation}\label{eq:design_space}
	Space = \prod\limits_{i=0}^{N-1} (O_i+1)
\end{equation}
where $O_i$ is the number of alternatives for the locking point $i$ plus the possibility of not locking the point. The size of the design space is thus proportional to the functional complexity of the region to be protected. The designers can apply more design-specific knowledge to restrict the analysis to specific code portions (i.e., critical sub-functions) or prune the total number of candidate points to speed-up the computation. The rest of the flow will operate only on the candidate points resulting from this analysis step.

\begin{figure}[t]
\includegraphics[width=\columnwidth]{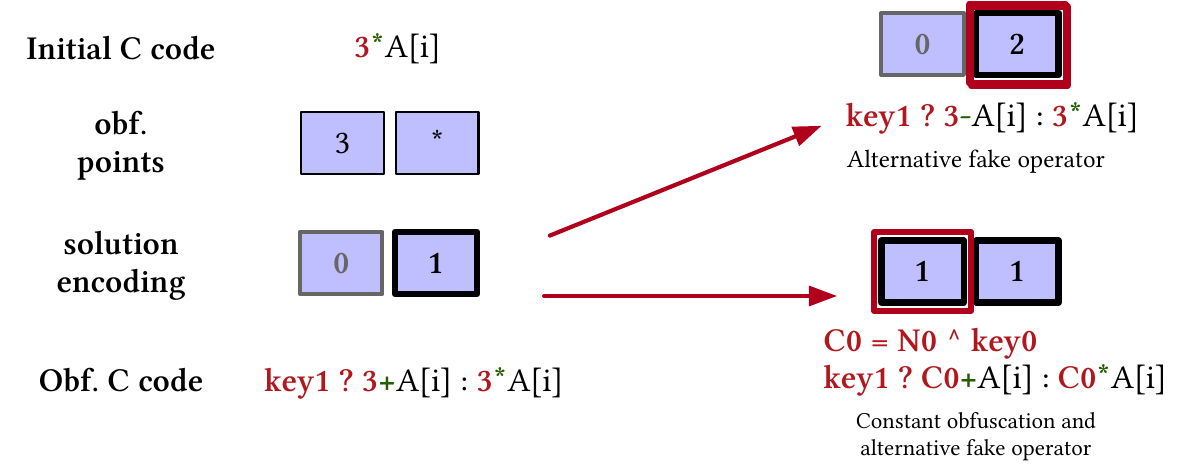}
\caption{Manipulation of locking solution to generate new ones.\vspace{-10pt}}\label{fig:mutation}
\end{figure}

\subsection{Exploration Phase and Security Assessment}\label{sec:dse}

In the {\em exploration} phase, we can use and compare different search methods to identify the best combination of locking techniques for the input C code. Our optimization framework can use any exploration method that is able to manipulate a vector of integers like in \Cref{fig:mutation}.

As a proof-of-concept, we implemented a standard GA with integer encoding. GAs maintain a population of $N$ alternative solutions (initialized randomly) and re-combine them to identify the best sub-set of locking techniques. 
Classic GA operators, like {\em random mutation} and {\em single-point crossover}, are applied with probability $P_m$ and $P_c$, respectively, to generate offspring solutions. At the end of each generation, all individuals are ranked based on the given security metric, checking for the best solution and passing the best individuals to the next generation. The procedure terminates when the solution is not improved for some generations or we reach the limit on the number of generations.

Each solution $s$ encodes locking transformations to be applied. First, we compute the number of key bits (see \Cref{eq:keys}) to determine if the solution is feasible. We then proceed with security assessment. We consider a threat model without an oracle. So we optimize the {\em differential entropy} (i.e., effects on the output values), which is a behavioral metric. We perform security evaluation on the locked C code by computing differential entropy  in \Cref{eq:corruptibility}. We apply locking techniques to the C code in the solution vector. This new code is compiled with the testbench and executed on test vectors for each alternative wrong key $w \in W$. The outputs are compared with golden outputs to compute differential entropy $H_s$. Exploration phase maximizes this value. If the designers want to trade-off more (security) metrics, they can use a linear combination of the corresponding values or perform multi-objective optimization~\cite{4556831}.

\subsection{Resource Evaluation and Selection}\label{sec:solution_eval}

Applying different locking techniques can lead to the same security level but with different overheads. Once we identify the solutions that optimize the given security metric, we evaluate their resource consumption. First, to increase the number of solutions to evaluate, we pass to the {\em selection} phase solutions whose security metric is within a pre-defined range from the best ones. We obtain more solutions with a minimal degradation (pre-defined by the designer) on the security metric. We perform commercial HLS on the locked C codes to obtain the corresponding locked RTLs. We rank these RTL designs according to the use of hardware resources, selecting the best as the final solution.

\section{Experimental Results}
\label{sec:results}

To validate our solution, we implemented a prototype in Python. We used {\tt pycparser} parser (ver.~2.19) for C manipulation (analysis and locking) and the DEAP framework (ver.~1.30)~\cite{DEAP_JMLR2012} for the GA-based DSE. Due to the stochastic nature of the GA, we averaged the results over 30 runs. 

\begin{table}[t]
\caption{Characterization of the benchmarks.}\label{tab:source}
\centering
\begin{tabular}{llrrrr} 
\toprule
\begin{tabular}[c]{@{}l@{}}\\\end{tabular} &       & \multicolumn{4}{c}{\bf Locking Points}   \\ 
\cmidrule(l){3-6}
{\bf Benchmark}                                  & {\bf Suite} & {\bf \#Ctrl.} & {\bf \#Op.} & {\bf \#Const.} & {\bf \#Bits}  \\ 
\midrule
\texttt{arf}                                        & Bambu \cite{bambu_dac_2021}     & 0       & 28    & 0        & 28      \\
\texttt{patricia}                                   & MiBench \cite{mibench_2001}     & 2       & 9     & 3        & 107     \\
\texttt{bubblesort}                                 & Bambu \cite{bambu_dac_2021}    & 0       & 11    & 4        & 139     \\
\texttt{crc}                                        & MiBench \cite{mibench_2001}   & 0       & 5     & 7        & 167     \\
\texttt{sha}                                        & MiBench \cite{mibench_2001}    & 0       & 76    & 40       & 1,356   \\
\texttt{adpcm}                                      & CHStone \cite{chstone_2008}     & 7       & 121   & 69       & 2,336   \\
\texttt{aes}                                       &  CHStone \cite{chstone_2008}     & 4       & 111   & 149      & 4,883   \\
\texttt{gsm}                                        & CHStone \cite{chstone_2008}    & 29      & 251   & 172      & 5,784   \\
\bottomrule
\end{tabular}
\end{table}

\begin{figure*}[!t]
\begin{subfigure}{0.49\textwidth}
\includegraphics[width=\textwidth]{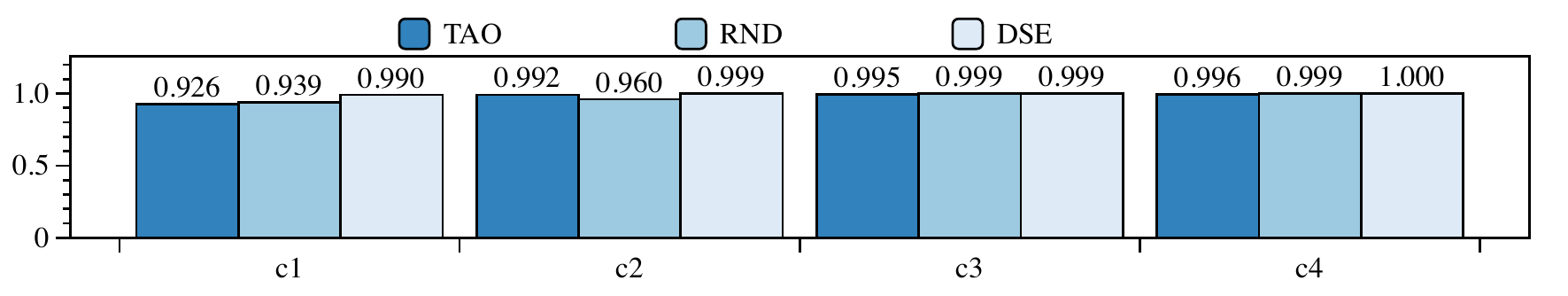}
\vspace{-18pt}\caption{\texttt{arf}}
\end{subfigure}\hfill
\begin{subfigure}{0.49\textwidth}
\includegraphics[width=\textwidth]{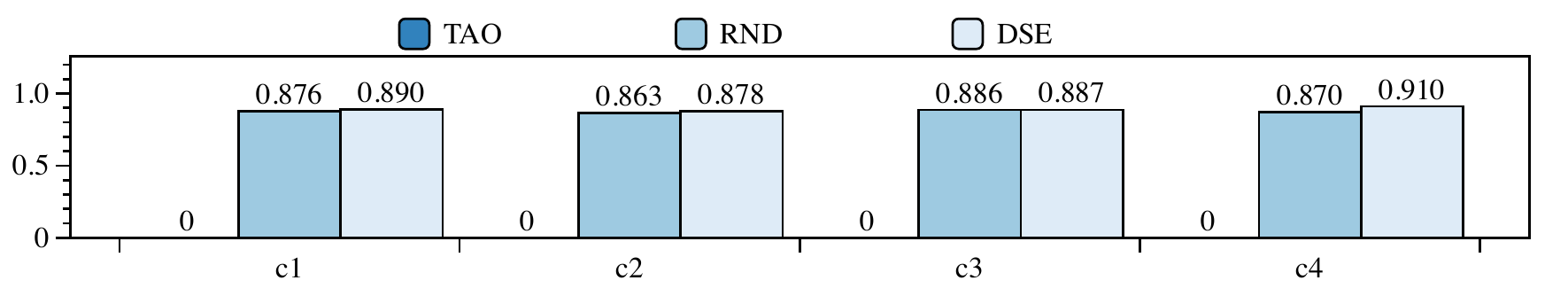}
\vspace{-18pt}\caption{\texttt{patricia}}
\end{subfigure}\\[4pt]

\begin{subfigure}{0.49\textwidth}
\includegraphics[width=\textwidth]{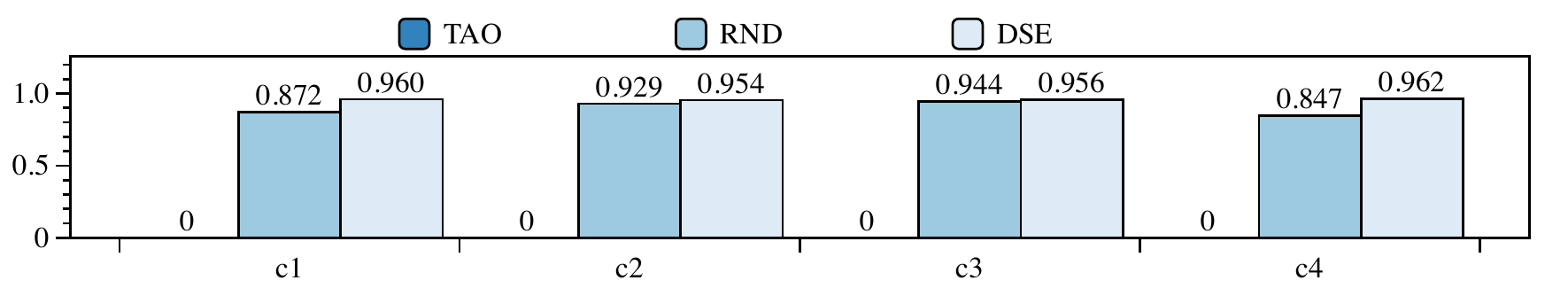}
\vspace{-18pt}\caption{\texttt{bubblesort}}
\end{subfigure}\hfill
\begin{subfigure}{0.49\textwidth}
\includegraphics[width=\textwidth]{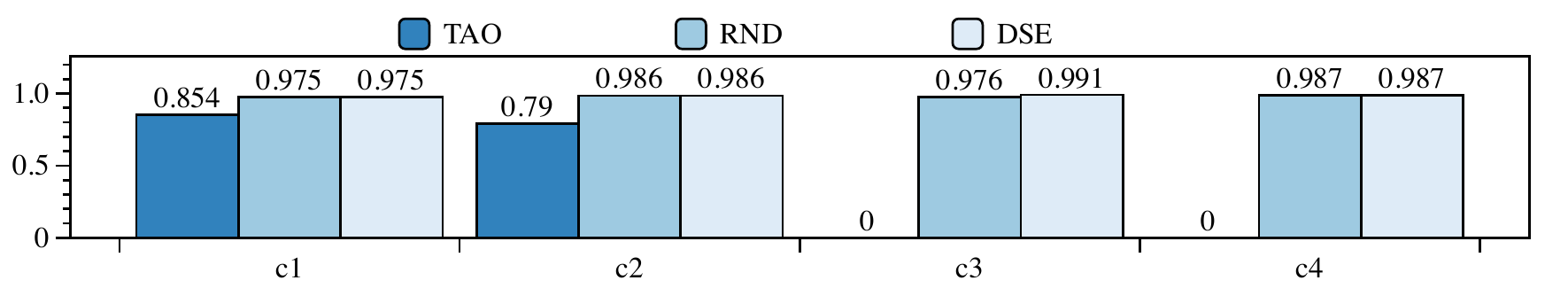}
\vspace{-18pt}\caption{\texttt{crc}}
\end{subfigure}\\[4pt]

\begin{subfigure}{0.49\textwidth}
\includegraphics[width=\textwidth]{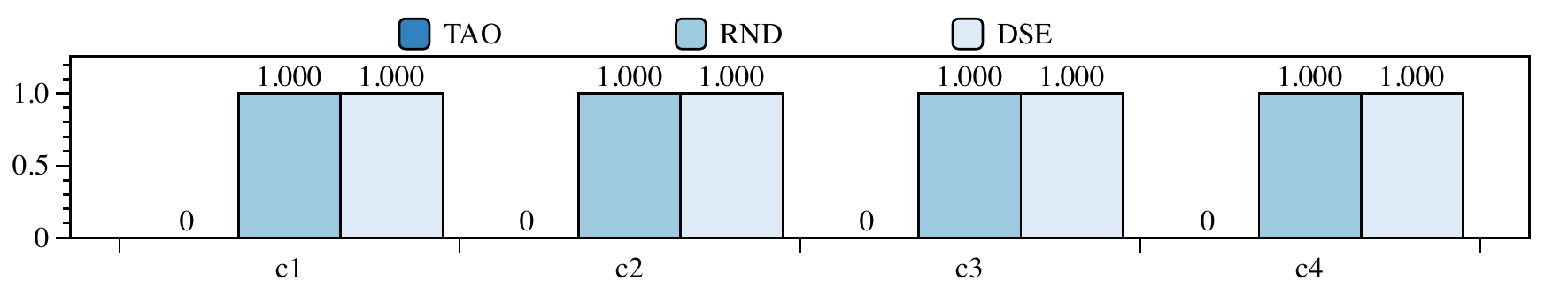}
\vspace{-18pt}\caption{\texttt{sha}}
\end{subfigure}\hfill
\begin{subfigure}{0.49\textwidth}
\includegraphics[width=\textwidth]{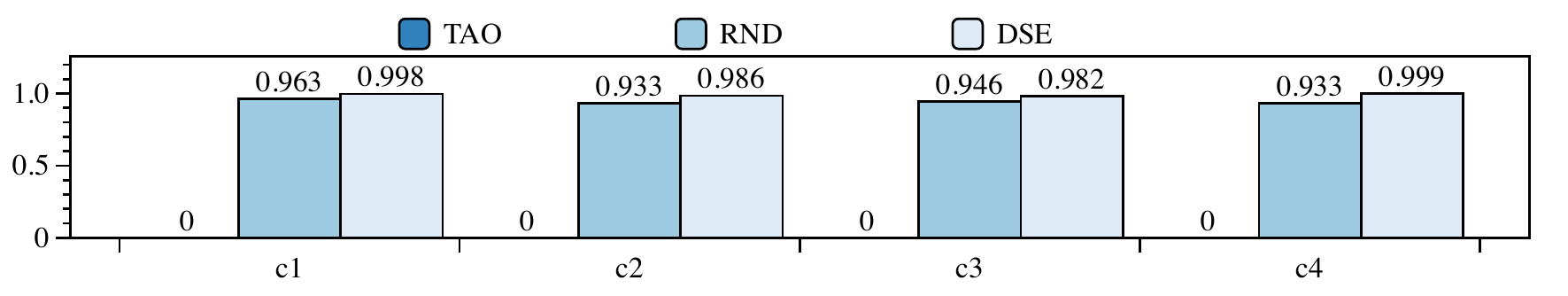}
\vspace{-18pt}\caption{\texttt{adpcm}}
\end{subfigure}\\[4pt]

\begin{subfigure}{0.48\textwidth}
\includegraphics[width=\textwidth]{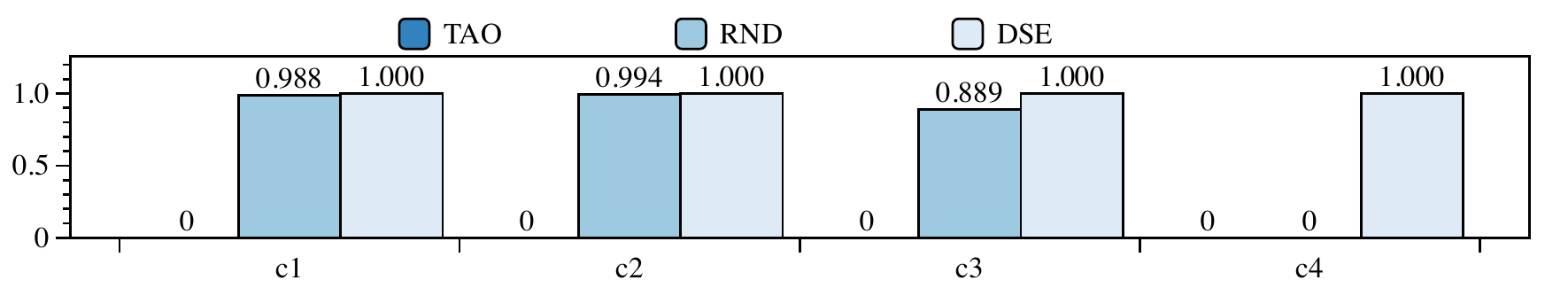}
\vspace{-18pt}\caption{\texttt{aes}}
\end{subfigure}\hfill
\begin{subfigure}{0.48\textwidth}
\includegraphics[width=\textwidth]{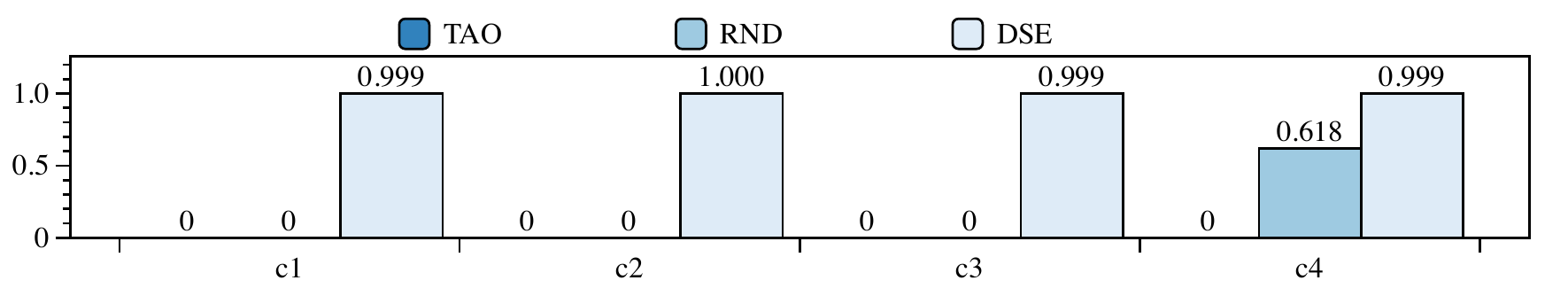}
\vspace{-18pt}\caption{\texttt{gsm}}
\end{subfigure}
\caption{Differential entropy comparison between our work (DSE) and topological locking (TAO) for different key budgets.}\label{fig:entropy_results}\vspace{10pt}
\end{figure*}

\begin{figure*}[!t]
\begin{subfigure}{0.48\textwidth}
\includegraphics[width=\textwidth]{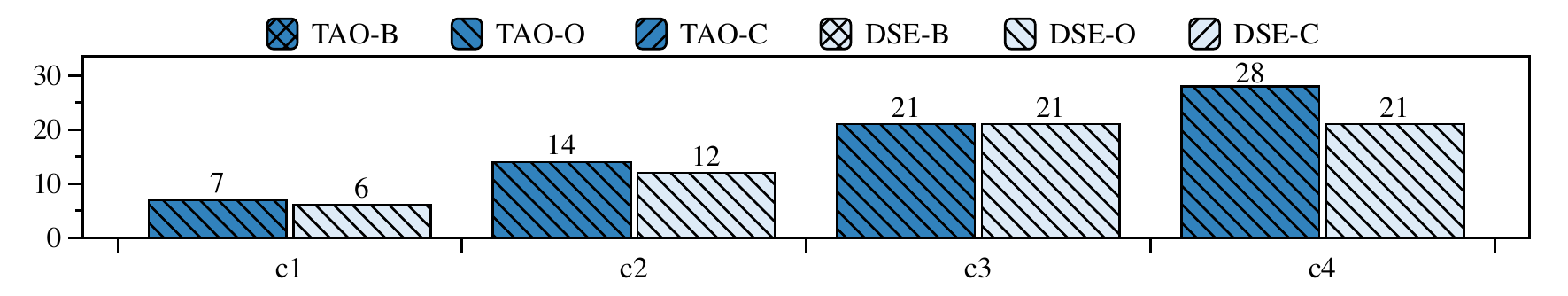}
\vspace{-18pt}\caption{\texttt{arf}}	
\end{subfigure}\hfill
\begin{subfigure}{0.48\textwidth}
\includegraphics[width=\textwidth]{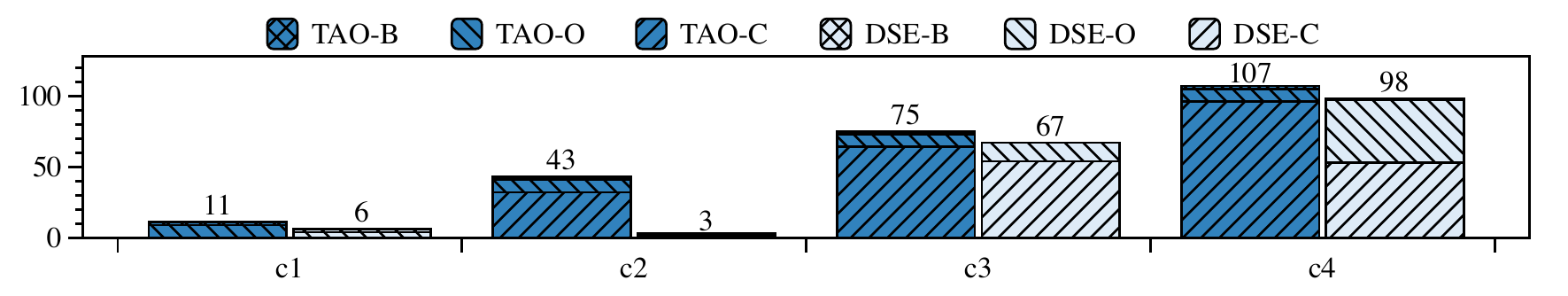}
\vspace{-18pt}\caption{\texttt{patricia}}	
\end{subfigure}\\[4pt]

\begin{subfigure}{0.48\textwidth}
\includegraphics[width=\textwidth]{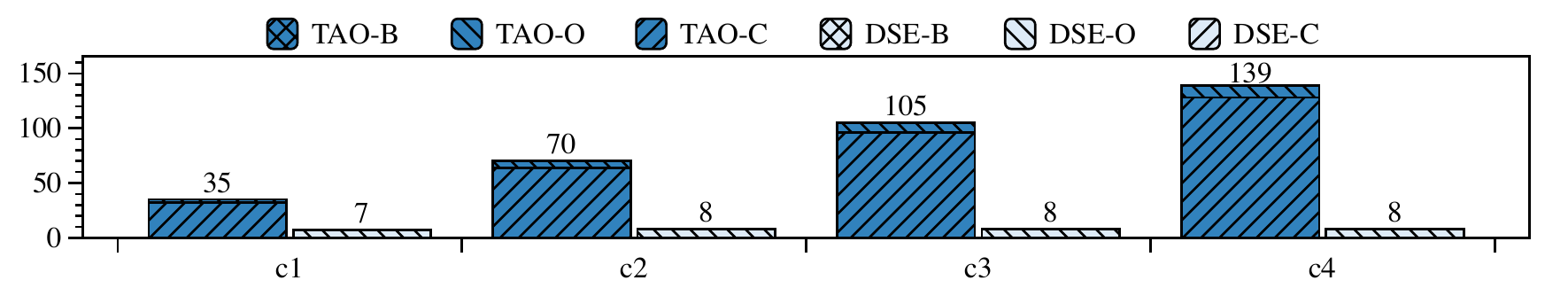}
\vspace{-18pt}\caption{\texttt{bubblesort}}	
\end{subfigure}\hfill
\begin{subfigure}{0.48\textwidth}
\includegraphics[width=\textwidth]{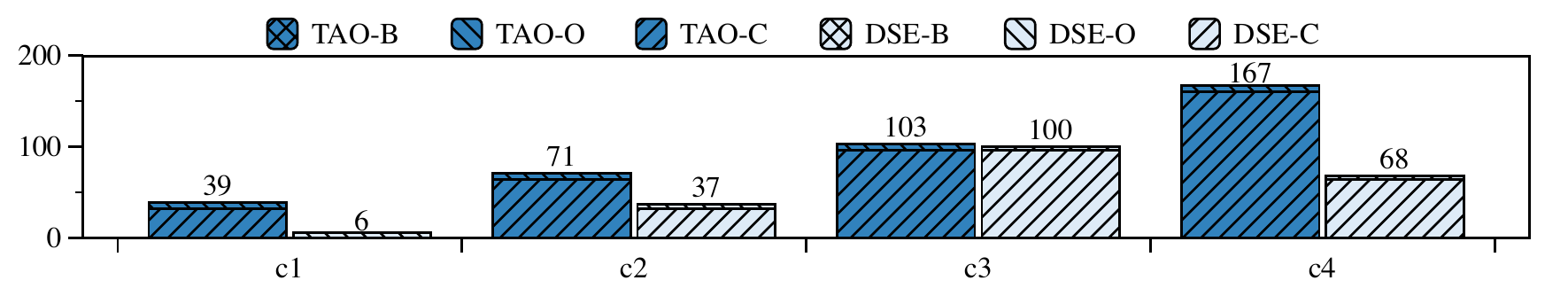}
\vspace{-18pt}\caption{\texttt{crc}}	
\end{subfigure}\\[4pt]

\begin{subfigure}{0.48\textwidth}
\includegraphics[width=\textwidth]{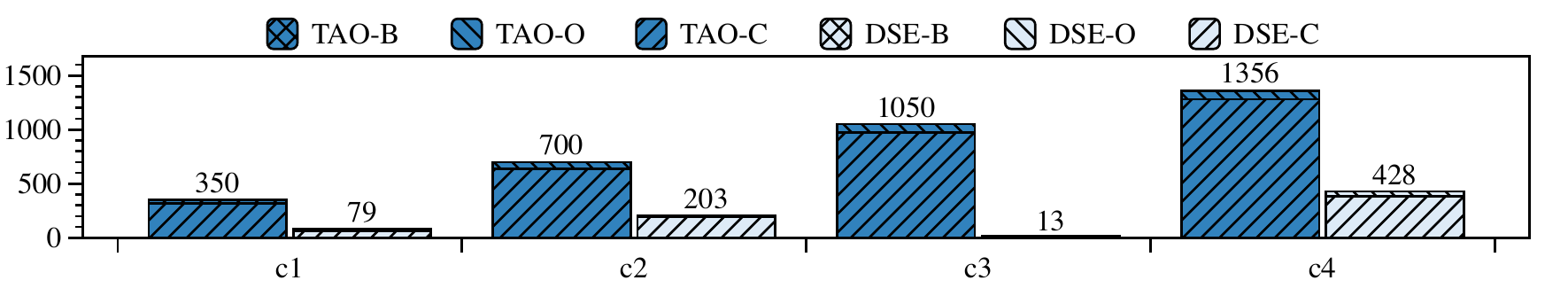}
\vspace{-18pt}\caption{\texttt{sha}}	
\end{subfigure}\hfill
\begin{subfigure}{0.48\textwidth}
\includegraphics[width=\textwidth]{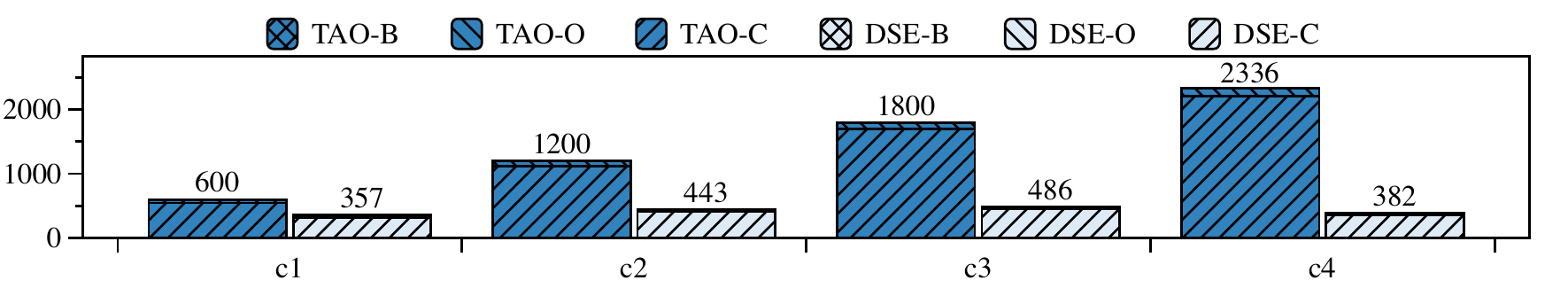}
\vspace{-18pt}\caption{\texttt{adpcm}}	
\end{subfigure}\\[4pt]

\begin{subfigure}{0.48\textwidth}
\includegraphics[width=\textwidth]{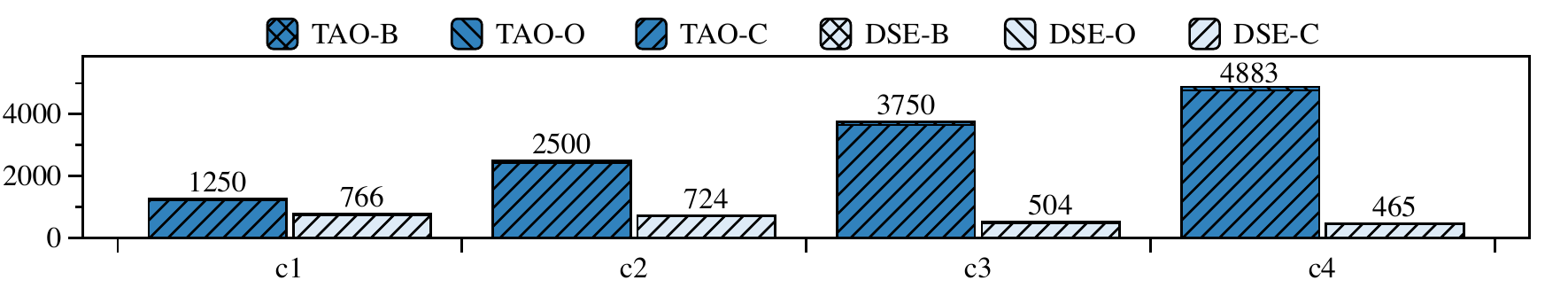}
\vspace{-18pt}\caption{\texttt{aes}}	
\end{subfigure}\hfill
\begin{subfigure}{0.48\textwidth}
\includegraphics[width=\textwidth]{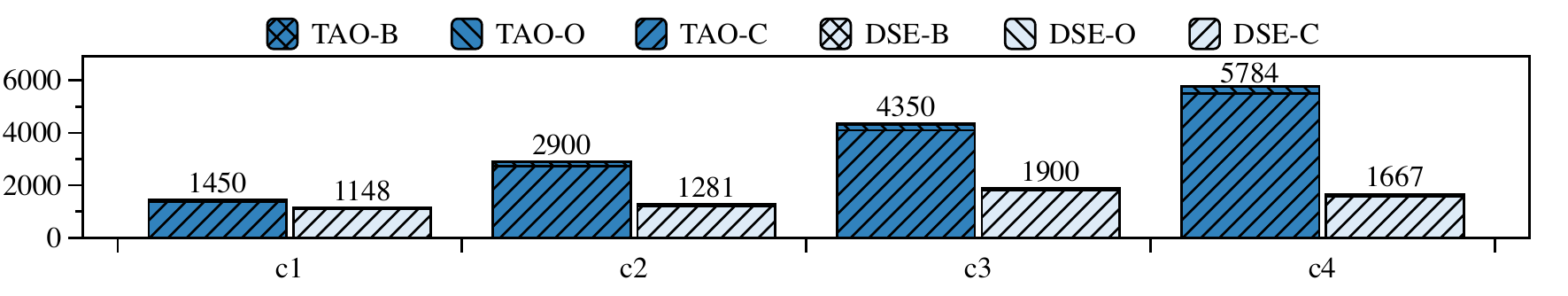}
\vspace{-18pt}\caption{\texttt{gsm}}	
\end{subfigure}
\caption{Number of locking bits used by our DSE and TAO topological locking for different key budgets. Each bar reports the number of bits used for locking constants (\texttt{*-C}), operations (\texttt{*-O}), and branches (\texttt{*-B}).}\label{fig:points_results}
\end{figure*}

\begin{table*}[t]
\caption{Hardware resources and corresponding overheads for topological locking (TAO) and our work (DSE).}\label{tab:overhead_results}
\centering
\setlength\tabcolsep{4pt} 
\vspace{-4pt}\begin{tabular}{lll rrrr rrrr rrrr rrrr} 
\toprule
\begin{tabular}[c]{@{}l@{}}\\\end{tabular} &   &    & \multicolumn{4}{c}{\texttt{arf}} & \multicolumn{4}{c}{\texttt{patricia}} & \multicolumn{4}{c}{\texttt{bubblesort}} & \multicolumn{4}{c}{\texttt{crc}}  \\
\cmidrule(l){4-7}\cmidrule(l){8-11}\cmidrule(l){12-15}\cmidrule(l){16-19}
\multicolumn{3}{c}{}                    & {\bf {\tt c1}} & {\bf {\tt c2}} & {\bf {\tt c3}} & {\bf {\tt c4}}     & {\bf {\tt c1}} & {\bf {\tt c2}} & {\bf {\tt c3}} & {\bf {\tt c4}}         & {\bf {\tt c1}} & {\bf {\tt c2}} & {\bf {\tt c3}} & {\bf {\tt c4}}            & {\bf {\tt c1}} & {\bf {\tt c2}} & {\bf {\tt c3}} & {\bf {\tt c4}}      \\ 
\midrule
\multirow{5}{*}{\rotatebox{90}{Resources}} & \multirow{5}{*}{\rotatebox{90}{Plain}}  & \#LUT   & \multicolumn{4}{c}{644} & \multicolumn{4}{c}{185}  &  \multicolumn{4}{c}{374} &  \multicolumn{4}{c}{285}             \\
                                            &   & \#FF       & \multicolumn{4}{c}{247}   &  \multicolumn{4}{c}{123}   &  \multicolumn{4}{c}{233}   & \multicolumn{4}{c}{264}          \\
                                            &   & \#DSP      &  \multicolumn{4}{c}{33}   &  \multicolumn{4}{c}{0}   &  \multicolumn{4}{c}{0}   & \multicolumn{4}{c}{0}     \\
                                            &   & \#BRAM     & \multicolumn{4}{c}{0}     &  \multicolumn{4}{c}{0}   &  \multicolumn{4}{c}{0}   & \multicolumn{4}{c}{0}      \\
                                            &   & Power [mW] &  \multicolumn{4}{c}{442}  &  \multicolumn{4}{c}{377} &  \multicolumn{4}{c}{328}   & \multicolumn{4}{c}{327}          \\ 
\midrule
\multirow{10}{*}{\rotatebox{90}{Overhead}} & \multirow{5}{*}{\rotatebox{90}{TAO}}  & \#LUT   &  +523  &  +817  &  +1387  &  +1602         & +2162   &  +1769  &  +1846  &  +1842              & +143   & +315   &  +498  &  +955                & +1552   & +1430   &  +1860  &  +2149         \\
                                            &   & \#FF    &  +285  &  +353  &  +586  &  +751         &  +1229  &  +932  & +937   &  +937              &  +61  &  +223  &  +329  & +534                 &  +751  &  +815  &  +831  &  +740         \\
                                            &   & \#DSP   &  -15  &  -12  &  -24  & -30          &  -  &  -  & -   & -               &  -  &   - &   - & -                 &  -  & -   &  -  &  -         \\
                                            &   & \#BRAM  &  -  &  -  &  -  &  -         &  -  &  -  & -   & -               & -   &  -  &   - &      -            & -   & -   &  -  & -      \\
                                            &   & Power [mW] &  -15  &  -6  & -14   & -14          & +12   &  +7  &  +1  &  +1              & -   &   +4 & +26  &  +12                & +12   & +13   &  +22  & +27          \\ 
\cmidrule(l){2-19}
& \multirow{5}{*}{\rotatebox{90}{Plain}}                          & \#LUT   &  +564  &  +1841  & +2232   &  +1982         &  +510  &  +144  & +263   & +112               &  +673  &  +688  &  +673  &  +267                &  +1441  & +730   & +1471   &  +1522                                                    \\
                                           &    & \#FF    &  +377  &  +1079  &  +1218  & +1324          &  +262  &  +87  & +123   & +37               &  +518  &  +518  &  +518  &  +223                &  +751  & +385   &  +847  & +783          \\
                                            &   & \#DSP   &  -15  &  -11  & -20   & -23          &  -  &  -  & -   &  -              & -   & -   &  -  & -                 &  -  &  -  & - &  -         \\
                                            &   & \#BRAM  &  -  & -   &  -  &  -         &  -  &  -  &  -  &  -              &  -  & -   & -   & -                 & -   & -   &  -  & -          \\
                                            &   & Power [mW] &  -11  &  -9  & -25 & -17           &  +3  &  +1  &  +2  & +1               & +7   &  +8  &  +8  & +4                 & +22   &  +11  & +22   & +22          \\
\bottomrule
\end{tabular}

\vspace{4pt}
\setlength\tabcolsep{2pt} 
\begin{tabular}{lll rrrr rrrr rrrr rrrr} 
\toprule
\begin{tabular}[c]{@{}l@{}}\\\end{tabular} &   &    & \multicolumn{4}{c}{\texttt{sha}} & \multicolumn{4}{c}{\texttt{adpcm}} & \multicolumn{4}{c}{\texttt{aes}} & \multicolumn{4}{c}{\texttt{gsm}}  \\
\cmidrule(l){4-7}\cmidrule(l){8-11}\cmidrule(l){12-15}\cmidrule(l){16-19}
\multicolumn{3}{c}{}                    & \multicolumn{1}{c}{\bf {\tt c1}} & \multicolumn{1}{c}{\bf {\tt c2}} & \multicolumn{1}{c}{\bf {\tt c3}} & \multicolumn{1}{c}{\bf {\tt c4}}     &  \multicolumn{1}{c}{\bf {\tt c1}} & \multicolumn{1}{c}{\bf {\tt c2}} & \multicolumn{1}{c}{\bf {\tt c3}} & \multicolumn{1}{c}{\bf {\tt c4}}         &  \multicolumn{1}{c}{\bf {\tt c1}} & \multicolumn{1}{c}{\bf {\tt c2}} & \multicolumn{1}{c}{\bf {\tt c3}} & \multicolumn{1}{c}{\bf {\tt c4}}            &  \multicolumn{1}{c}{\bf {\tt c1}} & \multicolumn{1}{c}{\bf {\tt c2}} & \multicolumn{1}{c}{\bf {\tt c3}} & \multicolumn{1}{c}{\bf {\tt c4}}      \\ 
\midrule
\multirow{5}{*}{\rotatebox{90}{Resources}\phantom{..}} & \multirow{5}{*}{\rotatebox{90}{Plain}\phantom{..}}& \#LUT   & \multicolumn{4}{c}{3017} & \multicolumn{4}{c}{6543}  &  \multicolumn{4}{c}{9641} &  \multicolumn{4}{c}{6594}             \\
                                            &   & \#FF       & \multicolumn{4}{c}{2660}   &  \multicolumn{4}{c}{4345}   &  \multicolumn{4}{c}{7903}   & \multicolumn{4}{c}{3574}          \\
                                            &   & \#DSP      &  \multicolumn{4}{c}{0}   &  \multicolumn{4}{c}{67}   &  \multicolumn{4}{c}{14}   & \multicolumn{4}{c}{49}     \\
                                            &   & \#BRAM     & \multicolumn{4}{c}{4}     &  \multicolumn{4}{c}{2}   &  \multicolumn{4}{c}{14}   & \multicolumn{4}{c}{8}      \\
                                            &   & Power [mW]\phantom{..} &  \multicolumn{4}{c}{442}  &  \multicolumn{4}{c}{377} &  \multicolumn{4}{c}{328}   & \multicolumn{4}{c}{327}          \\ 
\midrule
\multirow{10}{*}{\rotatebox{90}{Overhead}} & \multirow{5}{*}{\rotatebox{90}{TAO}}             
                        & \#LUT   &  +1885  &  +6990  &  +8917  & +8873       
                                &  +13777  &  +12436  &  +16102  &  +17166              
                                &  +5250  &  +10898  & +20047 & +10892              
                                &  +17221  &  +23023  &  +20235  & +21503          \\
                    &   & \#FF    &  +291  &  +2768  & +3825   & +4042         
                                &  +5956  &  +5784  &  +8583  & +9613            
                                &  -469  &  +2982  &  +2865  & +2989              
                                & +8730   &  +13206  &  +11923 & +12951          \\
                    &   & \#DSP   &  -  &  -  &  -  & -         
                                &  +19  &  -5  &  -2  & -15               
                                &   +3 &  +3  &  +3  & +4                 
                                &  -10  &  -48  &  -48  & -48          \\
                    &   & \#BRAM  &  -  &   - &  -  &   -       
                                &  -  &  -  &  -  &  -              
                                &  -  &  -  &  -  & -                 
                                &  +4  &  -2  &  -2  & -2          \\
                    &   & Power [mW] &  +219  &  +154  &  +208  &  +237         
                                & +18  &  +46  &  +56  &  +43              
                                &  +12  & +7   & +10   &  +10                
                                &  -18  &  +23  &  -26  &  -8          \\ 
\cmidrule(l){2-19}
& \multirow{5}{*}{\rotatebox{90}{DSE}}
                          & \#LUT   & +8992   & +4379   & +7403   & +3286
                                    &  +6302  & +7376   & +5357   & +1758
                                    &  +4142  & ++1430   & +155   &  +5651
                                    &  +15210  & +28175   & +21469  &  +35325       \\
                        &   & \#FF    &  +6415  &  +2906  & +5236   & +3286
                                    &  +1894  &  +1090  & +1418   & +751  
                                    &  +1786  &  +339  &  +254  &  +232  
                                    &  +8900  &  +14275  &  +9452 & +16621         \\
                        &   & \#DSP   &  +6  & +3   &  -  &  +3  
                                    &  +14  &  +28  &  -4  &  -12      
                                    &  +6  &  +4  &  -  &    +1   
                                    &  +49  & +55   & +18   & +42   \\
                        &   & \#BRAM  &  -  &  -  &  -  &  -  
                                    &  -  &  -  &  -  &  -
                                    &  +2  &  -2  &  +2  & -2
                                    & +2   &  +2  &   +4 & -   \\
                        &   & Power [mW] &  +84  & +199 & +143   &  +4  
                                    &  +7  &  +26  &  +7  & +47
                                    &  +3  &  +1  &  +2  & +1           
                                    &  +289  & +369   &  +72  & +241  \\
\bottomrule
\end{tabular}
\end{table*}

We selected eight benchmarks from the Bambu~\cite{bambu_dac_2021}, MiBench~\cite{mibench_2001}, and CHStone~\cite{chstone_2008} suites. The benchmarks have been selected because used for HLS-based locking~\cite{8465830} and already supported by the given HLS tool.
Table~\ref{tab:source} characterizes benchmarks in terms of locking points (branches, operations, and constants) and the total number of key bits required for complete locking. Benchmarks are ordered by increasing number of total key bits.

We configured the GA as follows. GA population has 300 individuals evolved for 1,000 generations or until the best fitness value does not improve for 10 consecutive generations. Crossover and mutation probabilities are set to $P_c = 0.5$ and $P_m = 0.2$. Single-element mutation probability is set to $P_l = 0.05$. The initial population is randomly created.
For each benchmark, we consider 100 input sets to evaluate the differential entropy. For each benchmark, we generate four keys of different length, namely \texttt{c1}, \texttt{c2}, \texttt{c3}, and \texttt{c4} to evaluate effect of key size  (25\%, 50\%, 75\% and 100\% of the required key bits). For each key, we generate 100 random variants that represent {\em wrong keys} for security evaluation~\cite{metrics2018}. 
We compare our solution with TAO~\cite{8465830}, a state-of-the-art behavioral locking technique that locks the elements  depth-first (i.e., \textit{topological locking}), and a random locking. For fair comparison, we re-implemented TAO in our framework. In TAO, the security metric is evaluated on the final solution, without security optimization and is identified as {\tt TAO} in our experiments. Random locking corresponds to the best individual in the initial GA population, i.e., the best solution among 300 alternatives.

\subsection{Security Metric Optimization}
\label{sec:results_security}

This paper does not aim at evaluating the security of the locking techniques, which is given (see \cite{pilato2021assure} for more information on the security guarantees), but aims at optimizing their use for a given security metric. For each benchmark $s$, we computed the theoretical maximum ${H}^*_s$ of the security metric $H_s$ and we normalized the values obtained with DSE and TAO. The \textit{perfect differential entropy} is thus equal to 1 but it can be impossible to be achieved for some benchmarks due to the nature of their algorithms and operations. 

\Cref{fig:entropy_results} compares the differential entropy (normalized with respect to the maximum value) of the state-of-the-art topological locking (TAO), the random solutions (RND), and our method (DSE). The results clearly show that topological locking fails to optimize the security metric. The analysis on the AST is not able to predict the effects on the outputs, leading in most of the cases to solutions with differential entropy equal to zero. These cases happen when the locked solutions invalidate the algorithms with fixed outputs (e.g., always equal to zero) or leading to time-outs (e.g., in case of infinite loops). Also, the points where these solutions are invalidated depends on how the algorithm is written. For example, in the \texttt{crc} benchmark, the topological locking invalidates the design only when the locking impacts the second half of the locking points. These invalid solutions are instead discarded by our exploration method that is always able to find solutions that are close to the optimal value. Random locking can achieve good solutions, but it lacks scalability. Indeed, when increasing the key budget (i.e., \texttt{c4}), it is more probable to select invalidating locking points. Our method is instead able to discard those points thanks to the exploration and recombination of alternatives.

\Cref{fig:points_results} shows the number of bits used for locking the solutions and the breakdown for the three techniques. First, results show that topological locking uses pre-defined number of bits proportional to the budget, while our DSE method number of bits independent of the budget (in many cases less than the limit). The number of bits depends on  optimizing the security metric rather than the budget. Constant locking has the most impact on differential entropy as it is selected most and uses most of the key bits. Branches are less used since manipulating control flow is likely to produce invalid designs. 

\subsection{Locking Overhead}
\label{sec:results_overhead}

We use Bambu an open-source HLS tool~\cite{bambu_fpl_2012} to generate RTL corresponding to plain and locked C codes. Bambu uses gcc 4.8 targeting a Xilinx Virtex-7 FPGA XC7VX690T with a clock period of 10ns. We targeted FPGAs since the ASIC backend of Bambu is only partially supported~\cite{bambu_dac_2021}. However, results are comparable between the technologies. Logic synthesis is done using Xilinx Vivado 2019.2. 

\Cref{tab:overhead_results} reports the characteristics of the designs obtained from the plain (unlocked) C codes and the overheads of TAO and DSE for different key budgets (\texttt{c1}-\texttt{c4}). 
We report look-up tables (LUT) and flip-flops (FF), along with DSP and BRAM elements. We also report total power consumption (in mW) of the synthesized accelerators.
Our DSE has a better use of resources than TAO. There are cases when LUT overhead is more, due to the complex alternative operators. Our method selects fake operators to optimize for security, while TAO uses a pre-defined alternative. DSP changes are minimal and limited to cases where fake operations are implemented as multipliers. BRAM elements are generally not affected because locking is not applied to memory elements. There are few cases where constant values cannot be converted into BRAM look-up tables, reducing the number of these elements and increasing logic. Power consumption is incremented proportionally to the additional logic. Using behavioral locking with different operators affects the HLS scheduling and liveness of temporary values. This impacts the number of registers and number of flip-flops. On the other hand, performing HLS and logic synthesis after behavioral locking has two positive effects. First, it allows us to reorganize the microarchitecture in a way that does not affect the total number of clock cycles since extra fake operations are executed in parallel to original ones. Second, since the extra logic is small compared to the original design, locked designs always meet the clock period, even when the plain design has a small positive slack. In both cases, the major effect is an increase of the area overhead to instantiate the fake operations or to recover the slack and meet the timing.

\begin{table}[t]
\caption{Additional results for  DSE: Equivalent solutions and number of generations.}\label{tab:dse}
\footnotesize
\centering
\begin{tabular}{@{}L{1.7cm}@{} @{}R{0.9cm}@{}R{0.9cm} @{}R{0.9cm}@{}R{0.9cm} @{}R{0.9cm}@{}R{0.9cm} @{}R{0.9cm}@{}R{0.9cm}@{}}
\toprule
         & \multicolumn{2}{c}{\bf {\tt c1}} & \multicolumn{2}{c}{\bf {\tt c2}} & \multicolumn{2}{c}{\bf {\tt c3}} & \multicolumn{2}{c}{\bf {\tt c4}}\\
\cmidrule(l){2-3}  \cmidrule(l){4-5}  \cmidrule(l){6-7}  \cmidrule(l){8-9} 
 {\bf Benchmark} & {\bf \#Sol.} & {\bf \#Gen.} & {\bf \#Sol.} & {\bf \#Gen.} & {\bf \#Sol.} & {\bf \#Gen.} & {\bf \#Sol.} & {\bf \#Gen.} \\ 
\midrule
\texttt{arf}	&	4	&	29	&	8	&	58	&	10	&	39	&	27	&	38	\\
\texttt{patricia}	&	2	&	22	&	2	&	14	&	4	&	12	&	2	&	45	\\
\texttt{bubblesort}	&	10	&	22	&	1	&	22	&	20	&	17	&	7	&	35	\\
\texttt{crc}	&	19	&	11	&	24	&	11	&	12	&	14	&	47	&	11	\\
\texttt{sha}	&	95	&	20	&	39	&	16	&	144	&	16	&	53	&	39	\\
\texttt{adpcm}	&	4	&	60	&	2	&	79	&	1	&	39	&	3	&	93	\\
\texttt{aes}	&	25	&	56	&	34	&	51	&	52	&	43	&	25	&	48	\\
\texttt{gsm}	&	1	&	73	&	2	&	62	&	8	&	64	&	2	&	77	\\
\bottomrule
\end{tabular}
\end{table}

\subsection{Design Space Exploration Performance}
\label{sec:dse_results}

\begin{figure}[t]
\begin{subfigure}{0.49\columnwidth}
\includegraphics[width=\columnwidth]{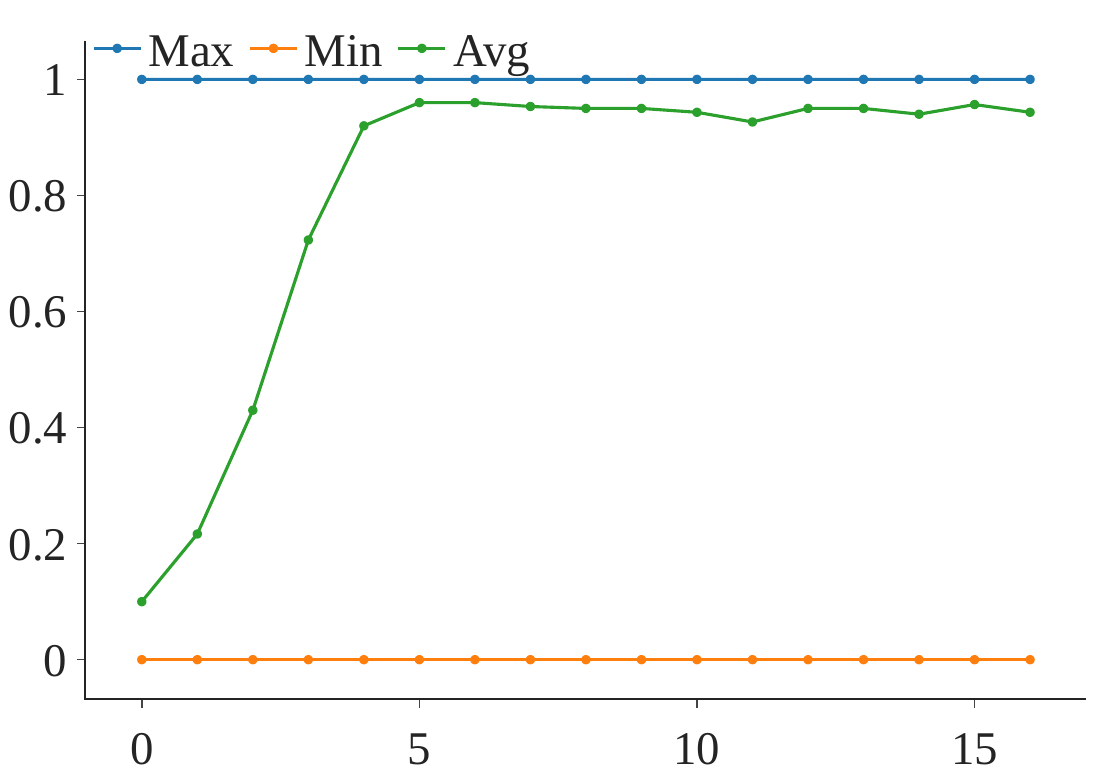}
\caption{\texttt{sha}}	
\end{subfigure}
\begin{subfigure}{0.49\columnwidth}
\includegraphics[width=\columnwidth]{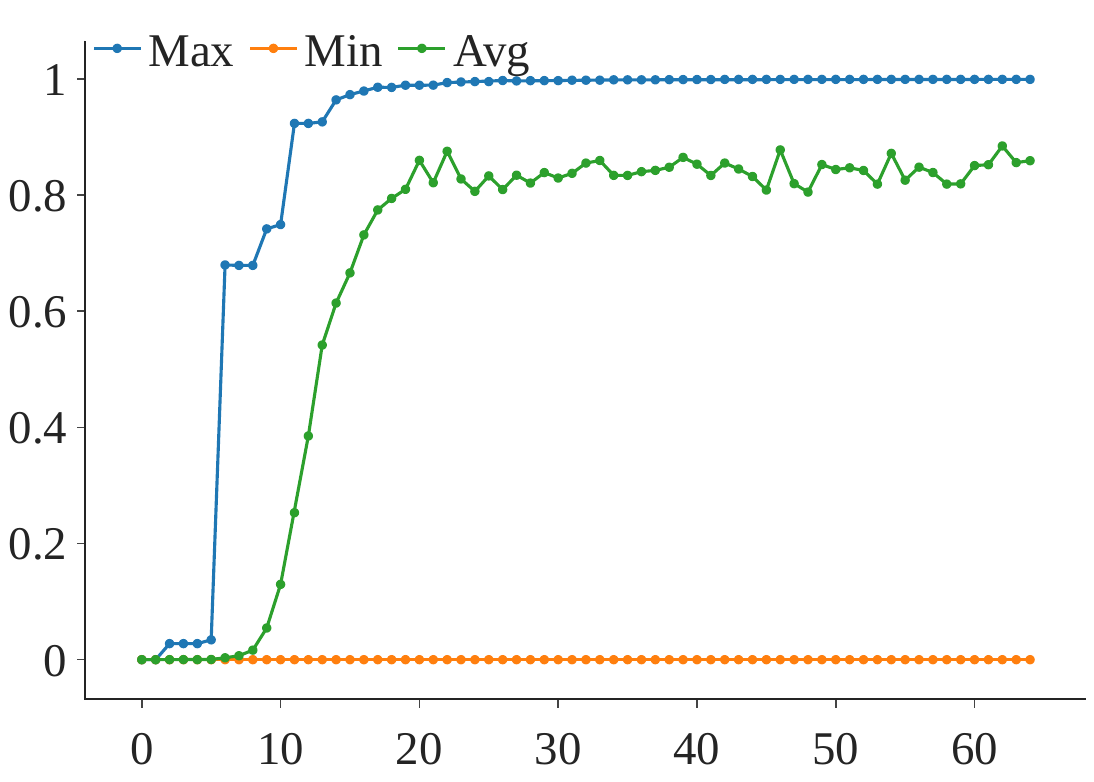}
\caption{\texttt{gsm}}	
\end{subfigure}
\caption{Evolution of the DSE exploration when providing 75\% of key budget (\texttt{c3}) for two representative benchmarks.}\label{fig:dse_results}
\end{figure}

\Cref{tab:dse}  shows the number of alternative solutions that are identified during exploration. Several solutions (up to 100) can reach a similar level of security. Remember that we maintain the solutions that are within an $\epsilon$-distance from the best one. Solutions that are resource hungry are filtered during  selection phase.
The design exploration that we perform is more expensive than single-run heuristics like TAO. While TAO completes the locking in few seconds in the worst case, our DSE engine requires many hours to complete (up to one day in the worst case). \Cref{tab:dse}  shows the number of generations required to converge to ``stable'' solutions. 
\Cref{fig:dse_results} shows the evolution of the DSE runs for two benchmarks (\texttt{sha} and \texttt{gsm}) when the key budget is 75\%. This is an interesting trade-off between a large design space and the constraint given by the number of key bits. 
The graphs show how the explorations progress towards better values for the security metrics. For \texttt{sha}, the best solutions are easy to find and the best value of the metric is almost immediate. The exploration terminates quickly. For the \texttt{gsm} benchmark, the DSE requires more generations to optimize the metric since the design is complex.  The average is sometimes decreasing  because the exploration phase introduces new individuals that are not necessarily better than the previous ones. However, ranking and selection procedures keep the best ones and use the worst ones to explore alternative regions of the design space. 
These results show that the method is robust and converges in about 20-30 generations. The execution time is affected mostly by time to compile the C code and time to evaluate security. Complex benchmarks require more time even if they have small code. They may lead to longer execution times when invalid solutions time out. The execution times are order of magnitude lower than RTL simulations. However, the improvements justify the extra time required by the designer to explore different solutions. Such exploration is performed only once during the design phase. 

\section{Related Work}

Several approaches have been proposed to protect IP cores from reverse engineering and IP theft, including {\em split manufacturing}, \textit{camouflaging}, {\em watermarking}, \textit{logic locking}. 
Split manufacturing  divides the IC design in two parts that are fabricated in separate foundries~\cite{6513707,8203796}. Camouflaging hides the Boolean function of a gate at the layout level~\cite{Cocchi2014}. Watermarking allows certifying ownership of IP by embedding a designer's signature into the design~\cite{AbdelHamid2004}. Logic locking makes the circuit function dependent on a key unknown to the foundry~\cite{8852678}. These techniques can apply at the transistor~\cite{8714856}, logic~\cite{Sengupta2018, 8203758}, or behavior~\cite{7100906} levels.

Locking modifies the behavior of a circuit, which does not produce the expected outputs until it is ``activated" with a secret key. It protects the circuit from illegal copies since the correct execution requires access not only to the IC layout but also the key. This approach protects against attackers with access to the design files~\cite{8852678}. Circuit locking can be performed at logic~\cite{roy2008epic, 7362173} or behavioral level~\cite{5401214, 5966342}. Protection techniques depend on the threat model assumed by the designer: if the attacker has access to an activated chip ({\em oracle}), locking must resist SAT attacks~\cite{Shamsi2017}. If no oracle is available, the attacker can only analyze the design files, which hardly reveal knowledge on the structure or function. The attacker can only apply random guesses. To scale to larger designs approaches raise the abstraction level such as applying locking during HLS~\cite{8465830,8942150}, even if they require tool modifications. Indeed, DSE at the RTL level has been shown to suffer poor performance because of the high number of locking points~\cite{collini_date_2022}. On the contrary, our method can integrate algorithm-level analysis and pruning steps to reduce the number of candidate locking points. They are not compatible with industrial EDA flows. Existing methods propose alternative techniques without explicitly optimizing security metrics~\cite{metrics2018}. We focus on identifying the best combination of techniques by analyzing the effects on the security metric.

Design space exploration has been recently applied to hardware IP protection~\cite{Sengupta2015,Wang2019,Wang2021}. In~\cite{Sengupta2015}, the DSE optimizes an area-delay function for IP watermarking. In~\cite{Wang2019}, the designers analyze the effects of restricted design spaces on obfuscated specifications. In~\cite{Wang2021}, locking is applied to selectively protect specific regions of the search space and not the hardware IP core. However, in all cases the security metric was not the primary optimization goal.

\section{Potential Framework Extensions}\label{sec:extensions}

We propose a solution to optimize differential entropy for behavioral locking in an oracle-less attack scenario. However, our framework can be extended in several directions.

\vspace{4pt}
\textbf{Locking techniques:} Designers can integrate new locking techniques. They must define the values for the alternatives for the technique and modify the analysis step to generate the elements in the solution vector. They can also develop further analysis and pruning steps to reduce the design space.

\vspace{4pt}
\textbf{DSE heuristics:} Design space exploration meta-heuristics are transparent to the locking techniques. Assuming that all techniques are orthogonal with each other, any common operator for design space exploration generates valid solutions  by manipulating the vector of integers representing the locking solution. as shown in \Cref{fig:mutation}. 

\vspace{4pt}
\textbf{Locking metrics:} Our framework can {\em protect the circuit against oracle-based attacks}, where the security metric is resilience to SAT-attacks. In this case, HLS must be performed already during security evaluation to create RTL designs on which SAT attacks are performed~\cite{blind}. Solutions that are broken (i.e., the key is recovered) can be marked as infeasible and discarded. To limit the execution time of the exploration, the designer can use the number of SAT clauses as a metric that corresponds to the complexity of SAT attacks. Eventually, the designer selects the solution that minimizes overhead among the ones that maximize the resilience to SAT.

\section{Conclusions}
Although HLS is popular, security constraints are not yet supported by commercial HLS tools. Countermeasures are applied to the code executing on the processors or manually implemented into IP blocks yielding suboptimal and even insecure designs. HLS should consider security side-by-side performance and cost~\cite{8114281}. Recent solutions are adding security awareness into HLS~\cite{8465830,8356053,7906738}.  They are not yet mature for industrial adoption. Our method optimizes IP cores with locking before HLS while limiting the overhead via design space exploration at the C level. Locked RTL is obtained by using any HLS tools. This is a pathway for behavior locking of industrial designs using commercial design flows.
The proposed locking maximizes a given security metric (i.e., differential entropy) by exploring the locking effects with a genetic algorithm. Results demonstrate that full locking is not necessary to maximize security. By selecting the locking points one can maximize security while limiting resource overhead. Operating at the C level makes our solution compatible with commercial HLS. Future research will work in two directions. To improve the framework, we will evaluate and compare alternative DSE techniques. To expand its application, we will apply it to new scenarios and security metrics.

\section*{Acknowledments}

The research is supported in part by NSF Award (\# 1526405), ONR Award (\# N00014-18-1-2058), NSF CAREER Awards (\# 1553419), the NYU Center for Cybersecurity (\url{cyber.nyu.edu}), and the NYUAD Center for Cybersecurity (\url{sites.nyuad.nyu.edu/ccs-ad}).

\bibliographystyle{IEEEtran}
\bibliography{main}

\begin{IEEEbiography}[{\includegraphics[width=1in,height=1.25in,clip,keepaspectratio]{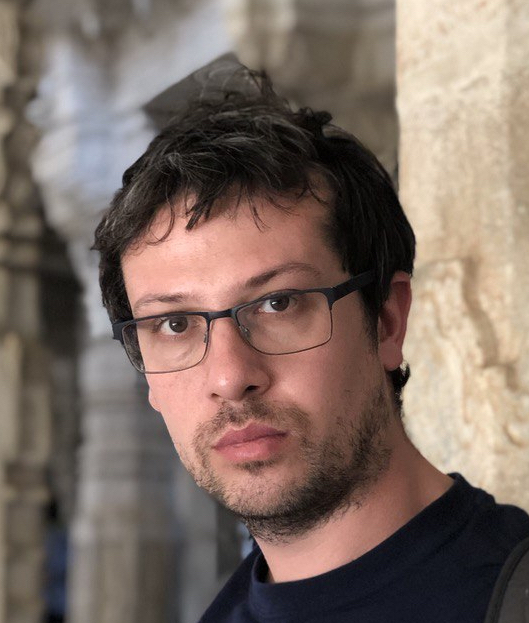}}]{Christian Pilato} 
is a Tenure-Track Assistant Professor at Politecnico di Milano. He was a Post-doc Research Scientist at Columbia University (2013-2016) and Università della Svizzera italiana (2016-2018). He was also a Visiting Researcher at New York University, TU Delft, and Chalmers University of Technology. He has a Ph.D. in Information Technology from Politecnico di Milano (2011). His research interests include high-level synthesis, reconfigurable systems and system-on-chip architectures, with emphasis on memory and security aspects. He served as program chair of EUC 2014 and is currently the program chair of ICCD 2022. He serves in the organizing and program committees of many conferences on EDA, CAD, embedded systems, and reconfigurable architectures (DAC, ICCAD, DATE, ASP-DAC, CASES, FPL, FPT, ICCD, etc.) He is a Senior Member of IEEE and ACM, and a Member of HiPEAC.
\end{IEEEbiography}

\begin{IEEEbiography}[{\includegraphics[width=1in,height=1.25in,clip,keepaspectratio]{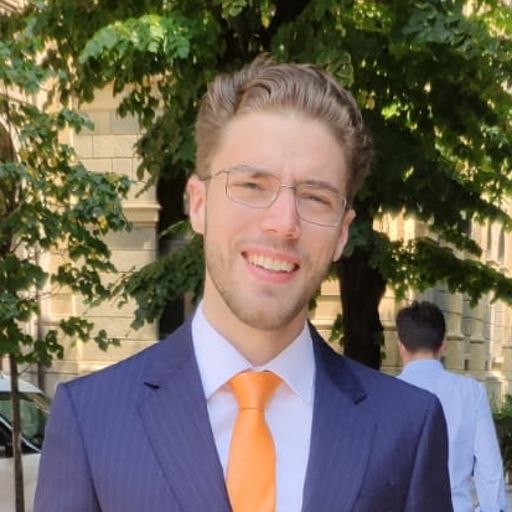}}]{Luca Collini} is a Ph.D. candidate at New York University (NYU) and an affiliated member of the NYU Center for Cyber Security (NYU CSS). He received his M.Sc. in Computer Science and Engineering from Politecnico di Milano in 2021 (summa cum laude) along with the Honourse Programme seal, which is a recognition for top-level students in the Computer Science program. He was also a research assistant at the same university until January 2022. His research interests include Electronics Design Automation, Intellectual Property (IP) protection and System-on-Chip (SoC) security, with particular emphasis on high-level methods to address these challenges.
\end{IEEEbiography}

\begin{IEEEbiography}[{\includegraphics[width=1in,height=1.25in,clip,keepaspectratio]{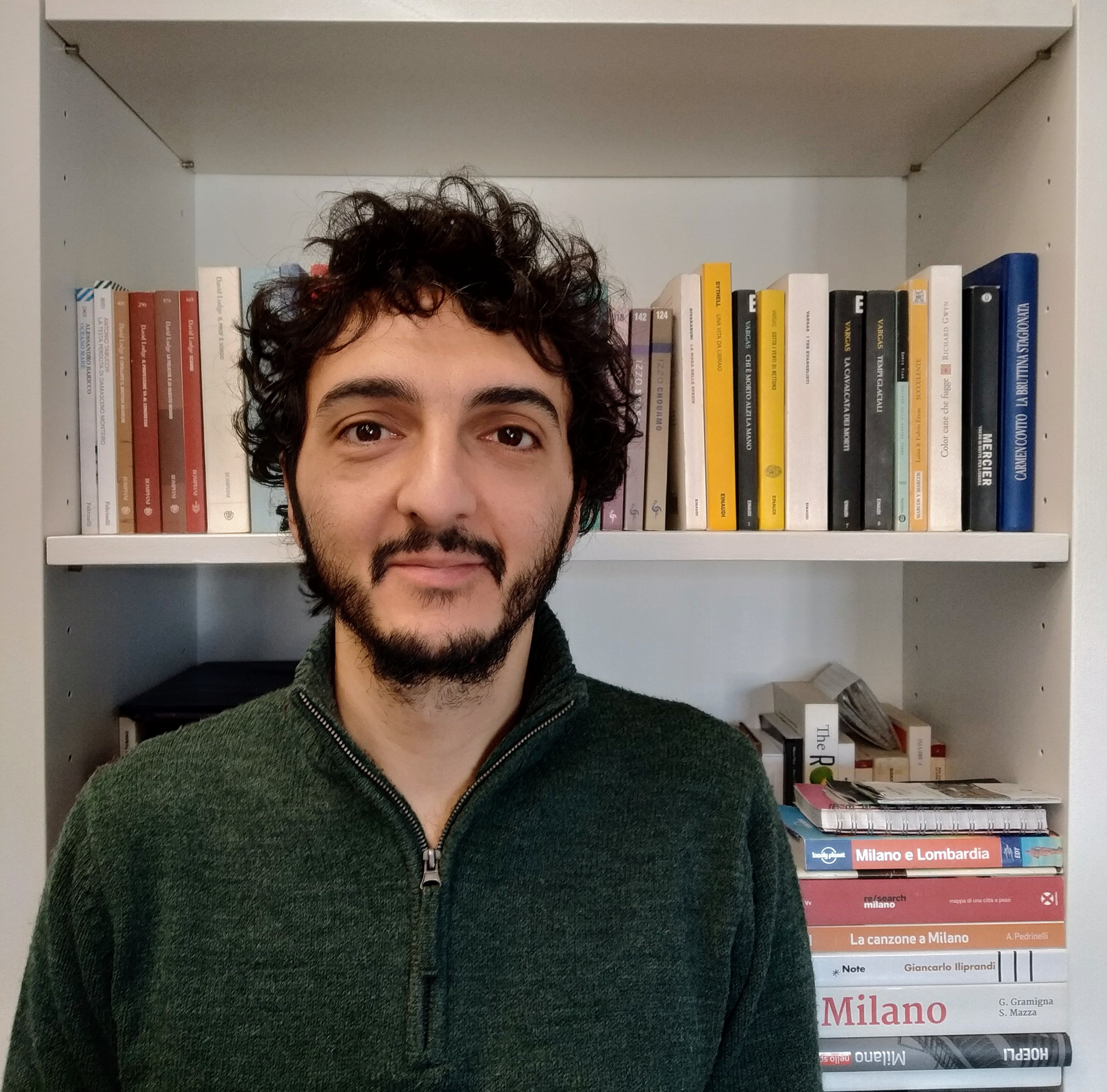}}]{Luca Cassano} is a Tenure-Track Assistant Professor at Politecnico di Milano, Italy. He received the B.S., M.S. and Ph.D. degrees in Computer Engineering from the University of Pisa, Italy. His research activity focuses on the definition of innovative techniques for fault simulation, testing, untestability analysis, diagnosis, and verification of fault tolerant and secure digital circuits and systems.
He served as program chair for DFTS 2021 and he is currently the program chair of DFTS 2022 and he serves in the organizing and program committees of several conferences on EDA, CAD and test (ETS, IOLTS, DDECS, DSD). He is associate editor of Integration, the VLSI Journal and of the Journal of Electronic Testing. With his Ph.D. thesis, titled ``Analysis and Test of the Effects of Single Event Upsets Affecting the Configuration Memory of SRAM-based FPGAs'', he won the European semifinals of the 2014 TTTC's E. J. McCluskey Doctoral Thesis Award.
\end{IEEEbiography}

\begin{IEEEbiography}
[{\includegraphics[width=1in,height=1.25in,clip,keepaspectratio]{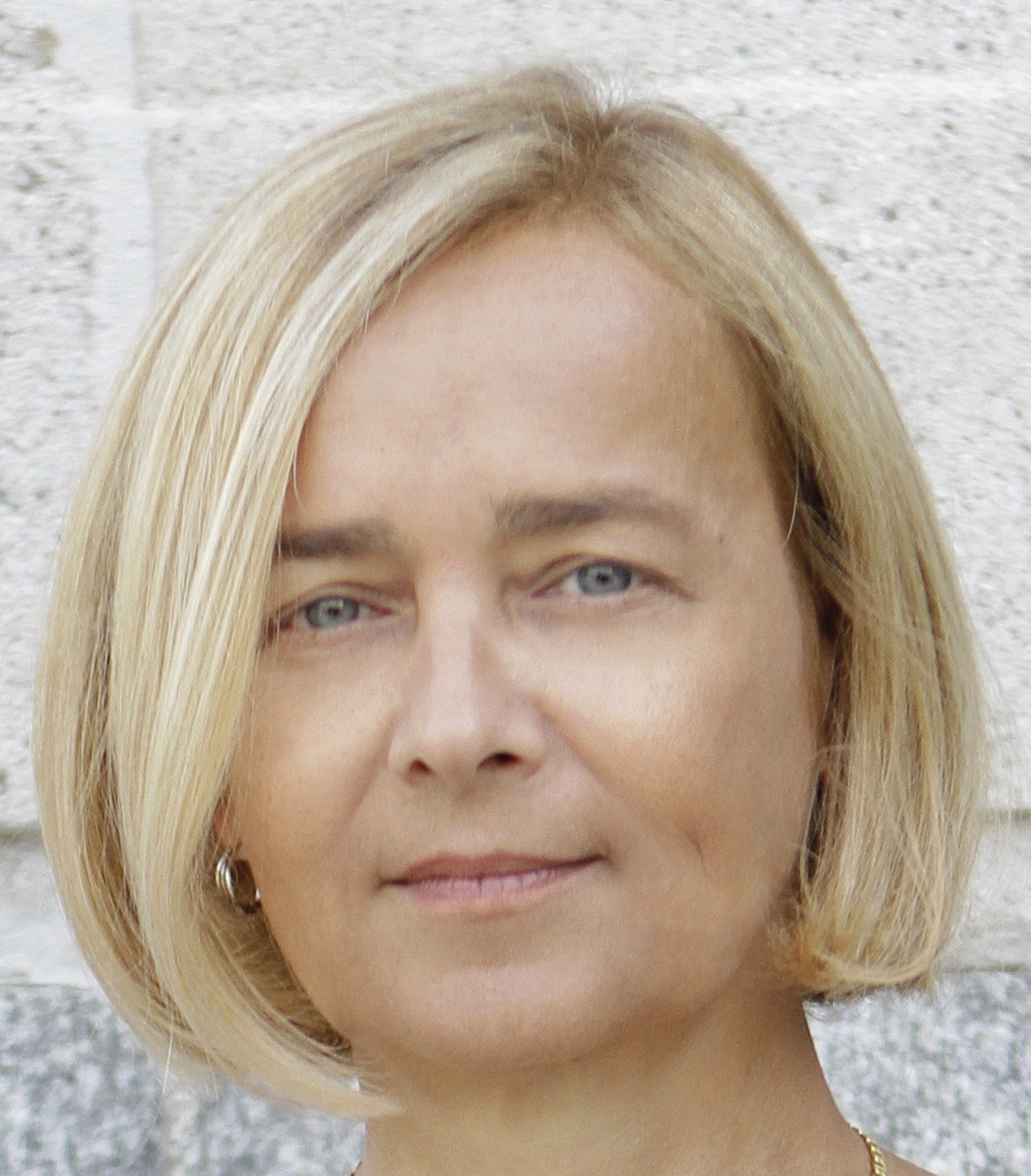}}]{Donatella Sciuto} 
received the  Laurea (Ms) in Electronic Engineering from Politecnico di Milano and the PhD in Electrical and Computer Engineering from the University of Colorado, Boulder, and the MBA from Bocconi University. 
She is currently the Executive Vice Rector of the Politecnico di Milano and Full Professor in Computer Science and Engineering. 
Her main research interests cover the methodologies for the design of embedded systems and multicore systems considering performance, power and security metrics. More recently she has been involved in managing and developing research projects in the area of smart cities and in the application of new ICT technologies to different application fields.  
She has published over 300 scientific papers.
She is a Fellow of IEEE for her contributions in embedded system design.  
She has served as Vice-President of Finance and then President of the IEEE Council of Electronic Design Automation from 2009 to 2013 and she serves in different capacities in IEEE Awards Committees, in scientific boards of IEEE journals and conferences. 
\end{IEEEbiography}

\begin{IEEEbiography}
[{\includegraphics[width=1in,height=1.25in,clip,keepaspectratio]{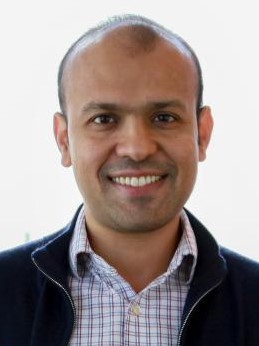}}]{Siddharth Garg} 
  received his Ph.D. degree in Electrical and Computer Engineering from 
Carnegie Mellon University in 2009, and a B.Tech. degree in Electrical 
Engineering from the Indian Institute of Technology Madras. He joined NYU in 
Fall 2014 as an Assistant Professor, and prior to that, was an Assistant 
Professor at the University of Waterloo from 2010-2014. His general research 
interests are in computer engineering, and more particularly in secure, reliable 
and energy-efficient computing. In 2016, Siddharth was listed in Popular 
Science Magazine's annual list of "Brilliant 10" researchers. Siddharth has 
received the NSF CAREER Award (2015), and paper awards at the IEEE Symposium on 
Security and Privacy (S\&P) 2016, USENIX Security Symposium 2013, at the 
Semiconductor Research Consortium TECHCON in 2010, and the International 
Symposium on Quality in Electronic Design (ISQED) in 2009. Siddharth also 
received the Angel G. Jordan Award from ECE department of Carnegie Mellon 
University for outstanding thesis contributions and service to the community. He 
serves on the technical program committee of several top conferences in the area 
of computer engineering and computer hardware, and has served as a reviewer for 
several IEEE and ACM journals.
\end{IEEEbiography}

\begin{IEEEbiography}[{\includegraphics[width=1in,height=1.25in,clip,keepaspectratio]{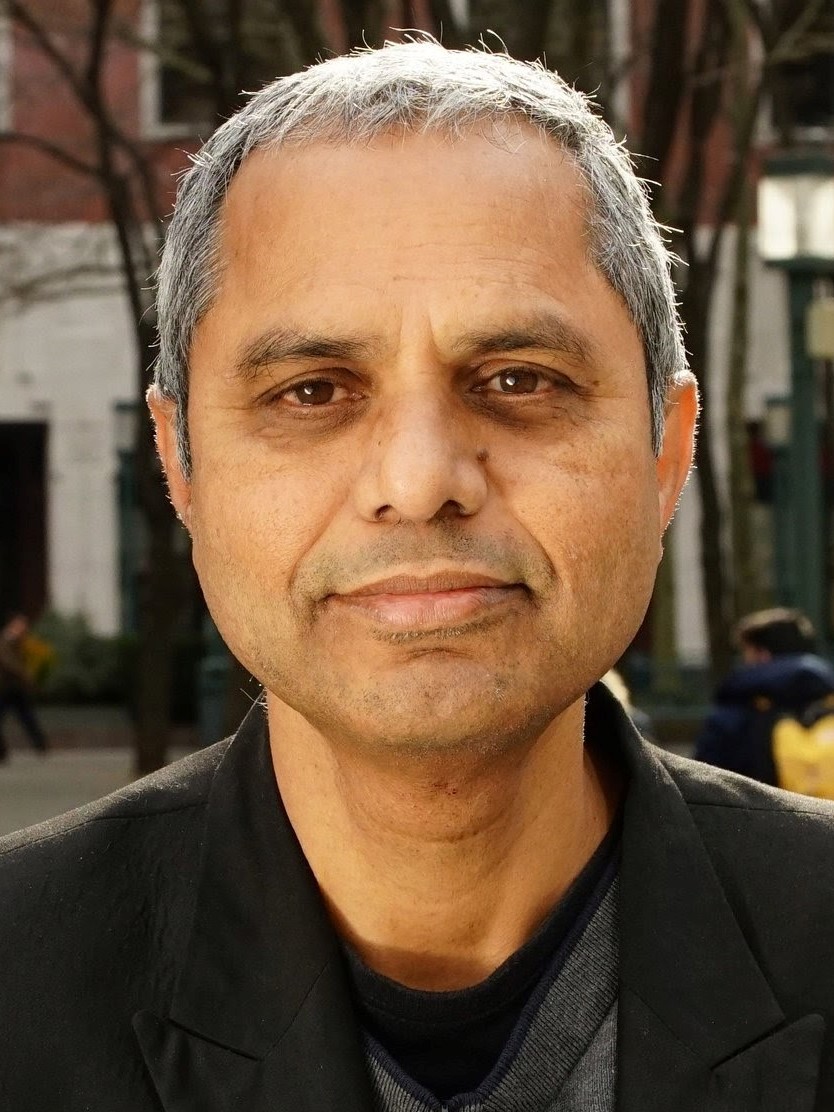}}]{Ramesh Karri} 
  is a Professor of ECE at New York University. He co-directs the NYU Center 
for Cyber Security (http://cyber.nyu.edu). He founded the Embedded Systems Challenge (https://csaw.engineering.nyu.edu/esc), the annual 
red team blue team event. He co-founded Trust-Hub (http://trust-hub.org). 
Ramesh Karri has a Ph.D. in Computer Science and 
Engineering, from the UC San Diego and a B.E in ECE from Andhra University. His 
research and education activities in hardware cybersecurity include trustworthy 
ICs; processors and cyber-physical systems; security-aware computer-aided 
design, test, verification, validation, and reliability; nano meets security; 
hardware security competitions, benchmarks, and metrics; biochip security; 
additive manufacturing security. He  published over 250 articles in leading 
journals and conference proceedings. Karri's work on hardware cybersecurity 
received best paper nominations (ICCD 2015 and DFTS 2015) and awards (ACM TODAES 
2018, ITC 2014, CCS 2013, DFTS 2013 and VLSI Design 2012). He received the 
Humboldt Fellowship and the NSF CAREER Award. He is the editor-in-chief of ACM JETC and serve(d)s  
on the editorial boards of IEEE and ACM Transactions (TIFS, TCAD, 
TODAES, ESL, D\&T, JETC). He was an IEEE Computer Society Distinguished 
Visitor (2013-2015). He served on the Executive Committee of the IEEE/ACM 
DAC leading the Security@DAC initiative (2014-2017). 
He served as program/general chair of conferences and serves on program committees. 
He is a Fellow of the IEEE for leadership and contributions to Trustworthy Hardware.

\end{IEEEbiography}
\end{document}